\documentclass[lettersize,journal]{IEEEtran}
\usepackage{amsmath,amsfonts}
\usepackage{algorithmic}
\usepackage{algorithm}
\usepackage{array}
\usepackage[caption=false,font=normalsize,labelfont=sf,textfont=sf]{subfig}
\usepackage{textcomp}
\usepackage{stfloats}
\usepackage{url}
\usepackage{verbatim}
\usepackage{graphicx}
\usepackage{cite}
\usepackage{bbm}
\usepackage{bm}

\newcommand{\indep}{\rotatebox[origin=c]{90}{$\models$}}
\newtheorem{theorem}{Theorem}[section]
\newtheorem{proposition}{Proposition}[section]
\newtheorem{definition}{Definition}[section]
\newtheorem{assumption}{Assumption}[section]

\begin{document}

\title{A Causal Perspective of Stock Prediction Models}

\author{Songci Xu, Qiangqiang Cheng, Chi-Guhn Lee
        % <-this % stops a space
\thanks{This paper was produced by the IEEE Publication Technology Group. They are in Piscataway, NJ.}% <-this % stops a space
\thanks{Manuscript received April 19, 2021; revised August 16, 2021.}}

\markboth{Journal of \LaTeX\ Class Files,~Vol.~14, No.~8, August~2021}%
{Shell \MakeLowercase{\textit{et al.}}: A Sample Article Using IEEEtran.cls for IEEE Journals}

\IEEEpubid{0000--0000/00\$00.00~\copyright~2021 IEEE}

\maketitle

\begin{abstract}
In the realm of stock prediction, machine learning models encounter considerable obstacles due to the inherent low signal-to-noise ratio and the nonstationary nature of financial markets. These challenges often result in spurious correlations and unstable predictive relationships, leading to poor performance of models when applied to out-of-sample (OOS) domains. To address these issues, we investigate \textit{Domain Generalization} techniques, with a particular focus on causal representation learning to improve a prediction model's generalizability to OOS domains. By leveraging multi-factor models from econometrics, we introduce a novel error bound that explicitly incorporates causal relationships. In addition, we present the connection between the proposed error bound and market nonstationarity. We also develop a \textit{Causal Discovery} technique to discover invariant feature representations, which effectively mitigates the proposed error bound, and the influence of spurious correlations on causal discovery is rigorously examined. Our theoretical findings are substantiated by numerical results, showcasing the effectiveness of our approach in enhancing the generalizability of stock prediction models.
\end{abstract}


\section{Introduction}
\IEEEPARstart{I}{n} the field of stock prediction, the application of machine learning models has seen increased emphasis. Although machine learning has achieved notable successes in various domains, predicting stock market trends remains particularly challenging due to the inherent low signal-to-noise ratio and the non-stationary nature of financial markets. The low signal-to-noise nature of the stock market often leads to spurious correlations, where random fluctuations are mistaken for meaningful patterns. Market nonstationarity, characterized by changing economic conditions and investor behaviors over time, further complicates the identification of stable predictive relationships. These characteristics frequently cause models, which perform well on historical data, to underperform when applied to out-of-sample (OOS) domains. Both aspects are widely recognized in the literature as enduring challenges for predictive modeling \cite{jiang2021applications, schmitt2013non}. 

In response to these challenges, it is logical to explore \textit{Domain Generalization} techniques, a research direction from the broader machine learning field aimed at enhancing model generalizability across different domains \cite{liu2021towards, wang2022generalizing}. Data augmentation is one of the cheapest examples of domain generalization technique. It has been effectively utilized to increase the quantity and diversity of datasets in numerous research areas, including computer vision (CV) and natural language processing (NLP). However, applying data augmentation techniques to financial time series is not straightforward. While there are some works focusing on developing data augmentation techniques for financial time-series \cite{aboussalah2022recursive, fons2020evaluating}, the field lacks a consensus on the best practices for augmenting heterogeneous raw inputs. It is still in the early stages of integrating domain generalization with financial prediction models. 

Causal learning represents another branch of domain generalization. This approach focuses on identifying the \textit{cause} of labels within the training domains, with the expectation that these causal relationships will persist in out-of-sample (OOS) domains \cite{christiansen2021causal, arjovsky2020out, mahajan2021domain}. Stock prediction problems are particularly well-suited to the causal perspective, as stock prices are extensively studied through multi-factor models in econometrics \cite{chincarini2006quantitative}. These multi-factor models are regression models designed to uncover the cause-and-effect relationships between a set of variables and a dependent variable. For instance, the Fama three-factor model \cite{fama1993common} examines the impact of the factors {\lq value\rq}, {\lq size\rq}, and {\lq market\rq} on stock returns. Unlike econometricians, who emphasize the economic interpretations of cause-and-effect relationships, we use the multi-factor model as an inductive bias for our domain generalization analysis. Specifically, we leverage the linear relationship between factors and stock returns imparted by these models, aiming to learn causal features that maintain a linear relationship with labels, regardless of the physical meanings of those features. While some studies \cite{lin2021deep, fang2020neural} have attempted to apply machine learning techniques to identify factors beneficial for predictive tasks, their approaches are typically heuristic and do not adopt a domain generalization perspective.  

In addition, traditional causal learning methods often rely on strong prior assumptions, such as the existence of a directed graph and invariant probabilities of labels conditional on causal factors \cite{zeng2021nonlinear, gilligan2022leveraging}. These assumptions can be excessively stringent for certain tasks, particularly in the chaotic financial market. Some works have relaxed these strong causal assumptions, positing that discovering \textit{invariance} is central to leveraging causality for generalization \cite{arjovsky2019invariant, arjovsky2020out}. For example, in computer vision (CV) problems, the focus is often on identifying invariant feature representations across domains rather than establishing causality among pixels. This principle is also illustrated by Newton's laws, which involve multiple physical variables on both sides of the equations, making it difficult to discern direct causality. What is crucial, however, is that these laws remain invariant across the universe. Similarly, our work follows this relaxation, aiming to identify invariant feature representations that maintain a linear relationship with stock returns, as characterized by multi-factor models. 
\IEEEpubidadjcol

In the theoretical analysis of domain generalization, it is common to study generalization error bounds. One approach involves examining the upper bound of the out-of-sample (OOS) prediction error, which helps identify aspects of training designs that contribute to low OOS prediction errors \cite{sicilia2023domain, ye2021towards}. Another approach focuses on the error bound of the distance between the average risk estimation across observable domains and the implicit true expectation of risks across all domains \cite{blanchard2021domain}. We adopt the former approach since considering expectations or averages across domains may not be suitable under market nonstationarity. 

A common issue with this type of error bound is the presence of an ideal joint error term in the upper bound, which is variable and intractable during the training process. The unwitting large values can completely ruin the model learning. Although some attempts have been made to replace the ideal joint error with more interpretable terms \cite{zhao2019learning}, most works barely assume that the ideal joint error remains small during the learning process. On the other hand, we found that, for the stock prediction problems, the ideal joint error is closely related to market nonstationarity. Therefore, it allows us to investigate how market nonstationarity affects the causal discovery.

Our contributions are threefold:
\begin{itemize}
    \item[i.]  We propose a novel error bound that explicitly accounts for causality induced by multi-factor models, ensuring that the ideal joint error remains constant once the training set is selected.
    \item[ii.] We introduce a theory linking the ideal joint error to market nonstationarity, making this term explainable and manageable if the training set comprises only recent time steps.
    \item[iii.] We develop the \textit{Causal Discovery} technique to learn invariant feature representations that reduces the proposed error bound, whose influence from spurious correlations are also rigorously analyzed.
\end{itemize}
This paper is structured as follows: In Section \ref{section: preliminaries}, we discuss the problem formulations, the issues of existing generalization error bounds from the literature, and the Structural Causal Models (SCM) induced by multi-factor models. The proposed error bound and the auxiliary analysis of the upper bound are elaborated in Section \ref{section: The Prediction Error Bound}. In Section \ref{section: Causal Discovery}, we rigorously discuss Causal Discovery. Numerical results from Section \ref{section: experiments} underpin our theoretical analysis and demonstrate the effectiveness of models trained using our causal discovery approach.

\section{Preliminaries, Settings $\&$ Discussions }
\label{section: preliminaries}
\subsection{Problem Formation}
The objective is to forecast a universe of stock returns on an out-of-sample (OOS) domain using a prediction model, denoted by $g{:}\, \mathbb{X} \rightarrow \mathbb{R}$, where $\mathbb{X}$ represents the input space. The model is trained on source domains. To define the formation of domains, consider the time steps $\{t_{-T-1}, t_{-T}, \ldots, t_{-1}, t_0\}$, where $t_{-T-1} < t_{-T} < t_{-T+1} < \cdots < t_{-1} < t_0$ and $T>0$ denotes the number of source domains. Information of daily granularity on the intervals $\{(t_{-i-1}, t_{-i}]\}_{i=1}^T$ is defined as $T$ source domains, and that on $(t_{-1}, t_0]$ is defined as the OOS domain. For the rest, the time step $t_{-i}$ indexes the domain on the interval $(t_{-i-1}, t_{-i}]$. Specifically, each $t_{-i}$ is considered a distinct domain and comprises $N_{t_{-i}}>0$ input-label pairs denoted by $\{(x^n_{t_{-i}}, r^n_{t_{-i}})\}_{n=1}^{N_{t_{-i}}}$, where $r^n_{t_{-i}}$ is the 1-day forward return of sample $n$ on domain $t_{-i}$. Here, $N_{t_{-i}}$ is defined as the number of stocks multiplied by the number of days in $(t_{-i-1}, t_{-i}]$. It is important to note that all domains do not necessarily contain the same number of days. For example, one can make each source domain contain only a couple of days in order to enlarge the number of domains within a fixed training set. On the other hand, a larger number of days in the OOS domain can extend the model's longevity, thereby reducing the frequency of model updates.

Following the tradition of supervised domain generalization, the learning process aims to minimize the average risk across all source domains. Mathematically, this is expressed as $\frac{1}{T}\sum_{i=1}^{T} \mathbf{err}_{t_{-i}}(g)$, where $\mathbf{err}_{t_{-i}}(g)$ is defined as $\int \ell(g(x), r) d\mathbb{P}_{t_{-i}}$. Here, $\ell : \mathbb{R} \times \mathbb{R} \rightarrow \mathbb{R}_+$ is a pre-defined distance function, and $\mathbb{P}_{t_{-i}}$ denotes the probability measure for domain $t_{-i}$. In this work, we use the Euclidean distance induced by the $\ell_2$-norm, denoted as $||\cdot||$. This approach leverages domain-wise insights and has been shown to outperform strategies that merely aggregate data across all domains into a single dataset \cite{wang2022generalizing, liu2021towards}. The ultimate goal is to minimize the OOS error on domain $t_0$, denoted as $\mathbf{err}_{t_0}(g) := \int \ell(g(x), r) d\mathbb{P}_{t_0}$. It is also important to note that the labels do not need to be the raw stock returns; for practical purposes, labels can be adjusted by demeaning or standardizing them to have zero mean and unit variance. This adjustment is justified by the sufficient condition that knowing the stocks' relative performances is adequate for profitable investment strategies. This paper focuses on standardized returns. 

Further, in order to measure the distances between probability measures, we define the Wasserstein distance of order 1, induced by $\ell_1$-norm $||\cdot||_1$, between $\mathbb{P}_{t_0}(x)$ and $\mathbb{P}_{t_{-i}}(x))$ as
\begin{equation}
\label{def: wass}
    \mathcal{W}^1(\mathbb{P}_{t_0}(x), \mathbb{P}_{t_{-i}}(x)): = \inf_{\pi\in\bm{\Pi}_{t_{-i}}}\int ||x_{t_0}- x_{t_{-i}}||_1 d\pi \text{\hspace{1em},}
\end{equation}
where $\bm{\Pi}_{t_{-i}}$ is the set of couplings between $\mathbb{P}_{t_{-i}}(x)$ and $\mathbb{P}_{t_0}(x)$ \cite{villani2009optimal}. 

\subsection{Generalization Error Bounds}
\label{section: generalization error bounds}
Our formulation of generalization error bounds is motivated by the error bounds in the domain adaptation context \cite{ben2010theory, johansson2019support}. In their context, let $\nu$ be a classifier from a hypothesis space $\mathcal{H}$ and $\phi$ a feature extractor, and consider the problem of transferring knowledge from a source domain to an OOS domain. We denote $t_{-1}$ to the source domain and $t_0$ to the OOS domain. Then, the risk bound is given by:
\begin{equation}
\label{errbound: DA}
    \textbf{err}_{t_0}(\nu \circ \phi) \leq \textbf{err}_{t_{-1}}(\nu \circ \phi) + \text{Distance}(\phi(X_{t_{-1}}), \phi(X_{t_0})) + \lambda^*(\phi),
\end{equation}
where $\text{Distance}(\cdot, \cdot)$ denotes any distance function between the source and OOS features induced by $\phi$, and $\lambda^*(\phi)$ is the ideal joint error defined as:
\begin{equation}
    \lambda^*(\phi) = \min_{\nu \in \mathcal{H}} \textbf{err}_{t_{-1}}(\nu \circ \phi) + \textbf{err}_{t_0}(\nu \circ \phi).
\end{equation}
With multiple source domains, the generalization error bounds are easily obtained by extending (\ref{errbound: DA}) to accommodate the convex combination of multiple source domains (e.g., \cite{ding2016incomplete}), aiming to transfer domain-invariant knowledge from multiple source domains to the OOS domain.

A common approach to reducing the OOS error involves jointly minimizing the source error and aligning the source and OOS features (i.e., minimizing the first two terms of (\ref{errbound: DA})) during the training process. However, counterexamples exist where the OOS error remains large even after applying adaptation techniques \cite{mahajan2021domain, zhao2019learning}. This issue arises because this approach implicitly assumes that $\lambda^*(\phi)$ is small and constant throughout the adaptation process, which is not generally true. Some works \cite{sicilia2023domain, zhao2019learning, arjovsky2020out} have identified the influence of $\lambda^*(\phi)$, arguing that a combination of small training errors and feature distances, alongside a large $\lambda^*(\phi)$, indicates negative transfer. Zhao et al. \cite{zhao2019learning} propose a lower bound on the OOS error, indicating that minimizing the first two terms of (\ref{errbound: DA}) can actually increase the OOS error if the marginal label distributions differ significantly between the source and OOS domains, thereby enlarging $\lambda^*(\phi)$ during the adaptation process. The unestimable and variable nature of $\lambda^*(\phi)$ during the learning process raises significant doubts about the effectiveness of popular domain adaptation techniques. 

\subsection{Converting Multi-Factor Models to Structural Causal Models}

\noindent According to cross-sectional multi-factor models \cite{chincarini2006quantitative}, the return of a sample $n$ in domain $t$ can be decomposed as:
\begin{equation}
    \begin{split}
    \label{eqn: multi-factor}
        r^n_t = z^{n,1}_tf^1_t + \cdots + z^{n,K}_t f^K_t + \epsilon^n_t,
    \end{split}
\end{equation}
where $r^n_t$ represents the return of sample $n$ in domain $t$, and $K >0$ signifies that $K$ factors are considered. These factor exposures, denoted by $z_t^{n, k}$, and factor premiums, denoted by $f^k_t$, are allowed to vary over domains, as indicated by the index $t$. The term $\epsilon^n_t$ refers to the error term, which in financial literature is often interpreted as the specific return of a stock. This return is influenced solely by events specific to the underlying company and cannot be explained by common factors. We leverage the inductive bias inherent in the multi-factor models by converting them to Structural Causal Models. 

To demonstrate, consider a straightforward yet insightful toy example of the SCM:
\begin{align*}
    Y_t &= Z_t + \mathcal{E}_t, \\
    X_t^a &= Y_t + \delta_t^a, \stepcounter{equation}\tag{\theequation}\label{SCM:toy} \\
    X_t &= (Z_t, X_t^a)^{\top}, \\
    \text{where } Z_t, \mathcal{E}_t, \delta_t^a &\sim \text{Normal}(0, \sigma_t^2).
\end{align*}
The subscript $t$ indexes the domain and $\sigma_t > 0$ varies over domains. In causal terms, $Z_t$ represents the \textit{cause} as it directly influences the labels, while $X_t^a$ is the \textit{effect}. The standard assumption in training data is that only the inputs $X_t$ and labels $Y_t$ are observable, while it is unknown which entries of $X_t$ correspond to the cause $Z_t$ or the effect $X^a_t$, and $\mathcal{E}_t$ remains unobservable. The effect often introduces spurious correlations as it correlates with the labels but lacks stable, generalizable properties. For instance, in the Cow \& Camel problem \cite{beery2018recognition, arjovsky2019invariant}, most cow images feature green backgrounds, and camel images often have desert backdrops. Consequently, a classifier trained in such a domain may erroneously associate green landscapes with cows and beige backgrounds with camels. Here, the animals represent the cause, and the landscapes are the effect, potentially leading to model confusion. As highlighted in \cite{arjovsky2020out, arjovsky2019invariant}, considering causality is pivotal due to the \textit{invariance} it imparts in models. Specifically, in predictive tasks, the primary concern is often not the causal relationships per se, but rather the identification of invariances that enable models to generalize to novel environments. Consider the example in (\ref{SCM:toy}): it can be demonstrated that the conditional expectation $\mathbb{E}_t[Y | Z]$ remains invariant over time, whereas $\mathbb{E}_t[Y | X^a]$ does not.\footnote{To illustrate, the time-varying regression coefficient for $\hat{Y}_t = \hat{a}_t X^a_t$ is given by $\hat{a}_t = 5\sigma_t^2$. This variation over time leads to a non-invariant $\mathbb{E}_t[Y | X^a]$.} This distinction is crucial, as identifying invariances helps to eliminate spurious correlations introduced by $X^a$ in this scenario. 

To bridge the multi-factor models with SCMs, the elements of the cross-sectional model in (\ref{eqn: multi-factor}) can be interpreted as instances of specific random variables:
\begin{equation}
    \label{eqn: RV}
    r^n_t \sim R_t, \hspace{1em} \epsilon^n_t \sim \mathcal{E}_t \hspace{0.5em}\text{and}\hspace{0.5em} z^{n,k}_t \sim Z^k_t \hspace{1em}\text{for $k = 1, \ldots, K$.}
\end{equation}
That is, the returns, specific returns, and factor exposures for these samples are realizations of the random variables $R_t$, $\mathcal{E}_t$, and $Z_t^k$ for $k = 1, \ldots, K$, respectively. Conventionally, the factor premiums within each domain $t$ are treated as deterministic, reflecting the static significance of each factor \cite{chincarini2006quantitative}. We thereby treat them as the \textit{causal coefficients}, denoted by ${F}^k_t$ for each factor $k$. The subscript $t$ indicates the domain-variant nature of these random variables. 

As to the stock market, the observable stock information, denoted by $X_{t}$, and the associated returns, denoted by $R_{t}$, are structurally defined by:
\begin{equation}
\label{SCM}
    \begin{split}
     & R_{t} = \mathbf{Z}_{t}^{\top}\mathbf{F}_{t} + \mathcal{E}_{t}, \\
     & X^a_{t} \dashleftarrow R_{t}, \\
     & X^a_t \dashrightarrow X_{t} \dashleftarrow \mathbf{Z}_{t}, \\
     & \text{where} \\
     & \mathbf{Z}_{t}, \mathbf{F}_{t}, \mathcal{E}_{t} \sim \mathbb{P}_{t}.
    \end{split}
\end{equation}

The first equation of (\ref{SCM}) is directly from (\ref{eqn: multi-factor}), with $\mathbf{Z}_{t} = (Z_t^1, \cdots, Z_t^K)$ and $\mathbf{F}_{t} = (F_t^1, \cdots, F_t^K)$. The model includes $K$ implicit factors for $K >0$, and we assume that $Z^k \indep Z^{k'}$ for $k \neq k'$ to ensure that each factor captures distinct information.\footnote{$\indep$ denotes independence.} Here, $\mathcal{E}_t$ signifies error term with $\mathbb{E}_t[\mathcal{E}] = 0$ and $\mathcal{E}_t \indep \mathbf{Z}_t$. In causality terms, $\mathbf{Z}_t$ is the \textit{cause}, while $X_t^a$ represents the \textit{non-causal factor}. The dashed arrows signify the variability of the relationship between $X_t^a$ and $R_t$ across environments. $X_t^a$ might be the \textit{effect} as in (\ref{SCM:toy}), potentially having spurious correlations with $R_t$ due to third-party confounders or being independent of $R_t$. The observable stock information $X_t$ encompasses both causal and non-causal factors. Contrary to the simplified scenario presented in the toy example (\ref{SCM:toy}), the relationships among $X_t$, $Z_t$, and $X_t^a$ are not explicitly stated in a general context. To represent this broader scope, we use dashed arrows in our diagrams, indicating that $\mathbb{P}_t(x|x^a, \mathbf{z})$ may vary across domains. 

\subsection{Transformed SCM}
\label{section: Transformed_SCMs}

When focusing on predicting standardized returns, the corresponding SCM can be precisely expressed as (\ref{SCM}) with $\mathbb{E}_t[R]=0$, $\mathbb{E}_t[R^2]=1$, and $\mathbb{E}_t[\mathbf{Z}] = \mathbf{0}$, where $\mathbf{0} \in \mathbb{R}^K$ is the zero vector (see \textit{Supplementary Materials}). While the causal mechanism of the toy SCM (\ref{SCM:toy}) is fixed across domains, the causal mechanism of (\ref{SCM}) changes over time, as depicted by the varying causal coefficients \(\mathbf{F}_t\). Generally, it is unnecessary to fully discover the cause \(\mathbf{Z}_t\) because the coefficients of most factors are volatile and challenging to predict over time. Instead, we concentrate on the stable subportion with more consistent causal coefficients. Specifically, for $0<\tilde{K}\leq K$, let $\bm{\Gamma}^{\tilde{K}}_t := \bm{\gamma}^{\tilde{K}} \bm{\Sigma}_t^{-1}$, where $\bm{\Sigma}_t$ is a diagonal matrix whose $k$th diagonal entry is $\sqrt{\mathbb{E}_t[(Z^k)^2]}$, and $\bm{\gamma}^{\tilde{K}} \in \mathbb{R}^{\tilde{K} \times K}$ is a time-invariant matrix such that $\bm{\gamma}^{\tilde{K}} = (\gamma_{ik})_{1 \leq i \leq \tilde{K}, \, 1 \leq k \leq K}$ and
\begin{equation}
\label{eqn: gamma}
    \begin{split}
            &\gamma_{ik} \gamma_{jk} = 0, \hspace{2em} \forall \, i \neq j, \, 1 \leq k \leq K,\\
            &\sum_{k=1}^K \gamma_{ik}^2 = 1, \hspace{2em} \forall \, 1 \leq i \leq \tilde{K}.
    \end{split}
\end{equation}

This subportion is characterized by \(\tilde{\mathbf{Z}}_{t} := \bm{\Gamma}^{\tilde{K}}_{t} \mathbf{Z}_{t} \in \mathbb{R}^{\tilde{K}}\), and the corresponding causal coefficients, denoted by \(\tilde{\mathbf{F}}_{t} := (\bm{\Gamma}^{\tilde{K}}_{t} \bm{\Gamma}^{\tilde{K} \top}_{t})^{-1} \bm{\Gamma}_{t}^{\tilde{K}} \mathbf{F}_t\), minimize the variance of the residuals \(\tilde{\mathcal{E}}_{t} := R_{t} - \tilde{\mathbf{Z}}_{t}^{\top} \tilde{\mathbf{F}}_{t}\) (see \textit{Supplementary Materials}). Note that $\bm{\gamma}^{\tilde{K}}$ characterizes the time-invariant economic interpretations for each entry of $\tilde{\mathbf{Z}}$, and $\bm{\Sigma}_t^{-1}$ serves to standardize the cause in each domain so the premiums become comparable across domains. Given this, the first line of (\ref{SCM}) can be transformed as:
\begin{equation}
\label{subportion}
    R_t = \tilde{\mathbf{Z}}_t^{\top} \tilde{\mathbf{F}}_t + \tilde{\mathcal{E}}_t.
\end{equation}
It can be checked that \(\mathbb{E}_t[\tilde{\mathbf{Z}}] = \mathbf{0}\) and \(\mathbb{E}_t[\tilde{\mathbf{Z}} \tilde{\mathbf{Z}}^{\top}] = \mathbf{I}\), where \(\mathbf{0} \in \mathbb{R}^{\tilde{K}}\) is the zero vector and \(\mathbf{I} \in \mathbb{R}^{\tilde{K} \times \tilde{K}}\) is the identity matrix.

After that, if the causal coefficients are invariant over domains, it is evident that:
\begin{equation}
\label{inv_tilde}
    \begin{split}
        &\tilde{\mathbf{F}}_{t_{-i}} = \tilde{\mathbf{F}}_{t_{-j}} \hspace{0.5em} \forall i, j \in \{0, \cdots, T\}\\
        &\implies \hspace{0.5em} \mathbb{E}_{t_{-i}}[R | \tilde{\mathbf{Z}}] = \mathbb{E}_{t_{-j}}[R | \tilde{\mathbf{Z}}] \hspace{0.5em} \forall i, j \in \{0, \cdots, T\}.
    \end{split}
\end{equation}
However, non-trivial invariant causal coefficients are unlikely to exist in reality.\footnote{Trivial invariant causal coefficients refer to coefficients that are all zeros over domains.} Instead, we allow the causal coefficients to deviate within a small range. Therefore, we give the following definition:\\
\begin{definition}
\label{def: generalizability}
    ($(\delta_{f}, \delta_{\epsilon})$-Generalizable.) Suppose we have source domains $t_{-1}, \cdots, t_{-T}$ and the OOS domain $t_0$. For $\delta_f,\delta_{\epsilon}>0$, we say the SCM of the form (\ref{SCM}) is $(\delta_{f},\delta_{\epsilon})$-generalizable if for all feasible transformations of the form (\ref{subportion}),
    \begin{equation}
    \label{eqn: generalizability}
    \begin{split}
        &\frac{1}{T}\sum_{i=1}^T ||\tilde{\mathbf{F}}_{t_{-i}} - \bm{\nu}^{*}|| \leq \delta_f \hspace{0.5em}\wedge\hspace{0.5em} \frac{1}{T}\sum_{i=1}^T \mathbb{E}_{t_{-i}}\left[|\tilde{\mathcal{E}}|\right] \leq \delta_{\epsilon}\\ &\implies \hspace{0.5em} ||\bm{\nu}^* - \tilde{\mathbf{F}}_{t_0}|| \leq \delta_{f}\hspace{1em} \wedge \hspace{0.5em} \mathbb{E}_{t_0}\left[|\tilde{\mathcal{E}}|\right] \leq \delta_{\epsilon},
    \end{split}
    \end{equation}
where $\bm{\nu}^* \in \arg\min_{\bm{\nu} \in \mathbb{R}^{\tilde{K}}} \frac{1}{T} \sum_{i=1}^T ||\tilde{\mathbf{F}}_{t_{-i}} - \bm{\nu}||$. Furthermore, $\tilde{\mathbf{Z}}$ is $(\delta_{f}, \delta_{\epsilon})$-generalizable causal features if its corresponding causal coefficients and residuals satisfy (\ref{eqn: generalizability}).\\
\end{definition}

This definition is motivated by the fact that if the in-sample causal coefficients all deviate within a small range, then the OOS coefficients are very likely to lie within that range as well. The \(\delta_{\epsilon}\) term in the definition signifies the need to avoid trivial causal features whose coefficients are constantly zero. A general pattern is that decreasing \(\delta_{\epsilon}\) will increase \(\delta_f\), and vice versa, while generalization is possible only if both $\delta_f$ and $\delta_{\epsilon}$ are small.

\section{The Prediction Error Bound}
\label{section: The Prediction Error Bound}

Let $\tilde{K} \in \mathbb{N}$ denote the dimensionality of the subportion corresponding to (\ref{subportion}), satisfying $0 < \tilde{K} \leq K$. We focus on the space of prediction models, denoted by $\Omega^g$, comprising functions of the form $g(x) = \phi(x)^{\top} \bm{\nu}$. Here, $\phi: \mathbb{X} \rightarrow \mathbb{R}^{\tilde{K}}$ is a function estimating causal features, and $\bm{\nu} \in \mathbb{R}^{\tilde{K}}$ is a time-invariant constant vector. Notably, if $\phi$ discovers $(\delta_f, \delta_{\epsilon})$-generalizable causal features with small $\delta_f, \delta_{\epsilon}> 0$, and $\bm{\nu}$ is selected as the mean of causal coefficients over source domains, then $g(x)$ maintains the generalizability towards the OOS domain $t_0$.

Our stock prediction problems differ from the settings associated with (\ref{errbound: DA}) in several ways. Firstly, we focus on regression problems rather than classification problems, eliminating the need to separately discuss the hypothesis function $\nu$ and the feature extractor $\phi$ in the error bound. Secondly, reducing the second term of (\ref{errbound: DA}) requires access to the marginal distributions of the OOS domain. However, we do not have access to the full marginal distribution until we pass the time point $t_0$. Given this, we consider the scenario where the OOS domain is entirely unknown.

Therefore, we propose a novel error bound in Theorem \ref{thm:err} where the second term of (\ref{errbound: DA}) is replaced by the Wasserstein distances, which measure joint probabilities involving features and labels. The Wasserstein distance imparts the causal perspective, which can be reduced if in-sample invariance is discovered and extended to the OOS domain. Also, the ideal joint error no longer varies with respect to the learning of $\phi$.  This modification allows us to incorporate the inductive bias of the financial market into the ideal joint error, making it more explainable and manageable.\\

\begin{theorem}
\label{thm:err}
Let $\lambda^*:=\min_{g\in \Omega^g}\frac{1}{T}\sum_{i=1}^T \textbf{err}_{t_{-i}}(g) + \textbf{err}_{t_0}(g)$. Then for all $g\in \Omega^g$, the OOS prediction error is bounded above as follows:
\begin{equation}
\label{errorbound}
    \begin{split}
        &\textbf{err}_{t_0}(g) \leq \frac{1}{T}\sum_{i=1}^T \textbf{err}_{t_{-i}}(g)\\ &+ \frac{1}{T}\sum_{i=1}^T \mathcal{W}^1(\mathbb{P}_{t_0}(g(x), r), \mathbb{P}_{t_{-i}}(g(x), r)) + 2\lambda^*,
    \end{split}
\end{equation}
where $\mathcal{W}^1(\mathbb{P}_{t_0}(g(x), r), \mathbb{P}_{t_{-i}}(g(x), r))$ is the Wasserstein distance of order 1 induced by $\ell_1$-norm.\\
\end{theorem}
%\noindent \textit{Proof}: See Appendix \ref{appendix: errorbound}. $\square$\\

The Wasserstein term encapsulates the causal perspective directly. It reflects the distances between source and OOS joint probabilities that capture the mechanism between the feature \(g(x)\) and the label \(r\).\footnote{The outputs induced by \(g\) can also be considered as features.} The following Proposition \ref{prop:wass} shows that if certain invariance is discovered, then this term can be greatly reduced:\\

\begin{proposition}
\label{prop:wass}
    Suppose $g\in\Omega^g$ and $\delta_f,\delta_{\epsilon}>0$. If $\phi$ discovers $(\delta_f, \delta_{\epsilon})$-generalizable causal features and $\mathbb{P}_{t_0}(\phi(x)) = \mathbb{P}_{t_{-i}}(\phi(x))$ for $i=1,\cdots, T$, then
    \begin{equation}
        \begin{split}
            \frac{1}{T}\sum_{i=1}^T \mathcal{W}^1(\mathbb{P}_{t_0}(g(x), r), \mathbb{P}_{t_{-i}}(g(x), r))\\
            \leq 2\delta_{\epsilon} + 2\sqrt{\tilde{K}}\delta_f.
        \end{split}
    \end{equation}\\
\end{proposition}
%\noindent \textit{Proof}: See Appendix \ref{appendix: propWass}. $\square$

The ideal joint error term $\lambda^*$ acts as a measure of how well-suited the source domains are for the OOS domain. Once the source domains are constructed, the ideal joint error is inherently fixed by its definition and remains unalterable, regardless of the techniques employed during the training process. When faced with a large $\lambda^*$, it is not feasible to ensure a small OOS error merely by minimizing the first two components of the error bound (\ref{errorbound}). Hence, it is imperative to preselect domains that are associated with low $\lambda^*$. Indeed, the joint error is closely related to the nonstationary nature of the stock market.

To study nonstationarity, we randomize the ideal joint error $\lambda^*$ by treating the OOS domain $t_0$ as a variable. It is reasonable to randomize $\lambda^*$ to account for multiple OOS domains: The equity curve of a portfolio constructed from consistent engagement in the stock market, and portfolio performance is usually calculated over numerous OOS domains. Formally, for $T>0$, we define $J_{T}:\mathbb{T}\rightarrow \mathbb{R}^+$ as $J_T(t_0):=\min_{g\in\Omega^g}\textbf{err}_{t_0}(g) + \frac{1}{T}\sum_{i=1}^T\textbf{err}_{t_{-i}}(g)$ where $\mathbb{T}$ is the set of all feasible domains. We inspect the probability of the ideal joint error exceeding a threshold, denoted by $\mathbb{P}(J_T>\tau)$ for $\tau>0$ being the threshold, and study where the minima are attained. To do so, we need to assume the existence of the minima:\\

\begin{assumption}
\label{assumption:NA}
    $\forall T>0$ and $\forall \tau>0$, the minima of the set $\{\mathbb{P}(J_{T'}>\tau) \mid T' \geq T\}$ exist.\\
\end{assumption}

Numerical simulations demonstrate that global minima are generally attained at smaller horizon in the training set for all $\tau > 0$ (see Section \ref{section: experiments}). This suggests that investors should consider selecting the most recent short time horizons to construct the source domains. However, the inherently low signal-to-noise ratio in stock market data complicates this approach. A small number of training samples can significantly hinder the learning process in this era, where models are generally overparameterized.

On the other hand, numerical results also suggest that the second minimum is achievable by slightly enlarging the training horizon from where the global minima are attained. Similarly, the third, fourth, fifth, and subsequent minima are also achievable by further enlarging the training horizon. To sum up, the $i$th minimum appears to smoothly depend on the length of training horizon, indicating that increasing the training horizon enlarges the sample size while gradually increasing the risk of a larger prediction error bound. As a result, investors face a critical trade-off between minimizing the effects of market nonstationarity and maintaining an adequate signal-to-noise ratio when learning prediction models.

Other than the large error bound, another influence from a large value of $\lambda^*$ is depicted by Proposition \ref{prop: NA_lbd}, elucidating two possible consequences: either (i) there may not exist a $\phi$ such that $\phi(X)$ approximates $\tilde{\mathbf{Z}}$ properly, or (ii) the corresponding SCM may not be $(\delta_f, \delta_{\epsilon})$-generalizable with small $\delta_f$ and $\delta_{\epsilon}$. Specifically, consider feasible $\bm{\gamma}^{\tilde{K}}$ that satisfy (\ref{eqn: gamma}) and result in the transformed SCM (\ref{subportion}) where $\tilde{\mathbf{Z}}$, $\tilde{\mathbf{F}}$, and $\tilde{\mathcal{E}}$ are dependent on $\bm{\gamma}^{\tilde{K}}$. The proposition is then stated as follows:\\

\begin{proposition}
\label{prop: NA_lbd}
Let $\delta_f(\bm{\gamma^{\tilde{K}}}, \bm{\nu}):= \frac{1}{T}\sum_{i=1}^{T}||\tilde{\mathbf{F}}_{t_{-i}} - \bm{\nu}|| + ||\tilde{\mathbf{F}}_{t_0} - \bm{\nu}||$, $\delta_{\epsilon}(\bm{\gamma}^{\tilde{K}}):= \frac{1}{T}\sum_{i=1}^T \mathbb{E}_{t_{-i}}\left[|\tilde{\mathcal{E}}|\right] + \mathbb{E}_{t_0}\left[|\tilde{\mathcal{E}}|\right]$, and $\delta_{z}(\phi, \bm{\gamma}^{\tilde{K}}):=\frac{1}{T}\sum_{i=1}^T\mathbb{E}_{t_{-i}}\left[ ||\phi(X) - \tilde{\mathbf{Z}}||\right] + \mathbb{E}_{t_{0}}\left[ ||\phi(X) - \tilde{\mathbf{Z}}||\right]$. Then,
\begin{equation}
\label{eqn:conseq}
    \lambda^* \leq \min_{\bm{\gamma}^{\tilde{K}}, \phi, \bm{\nu}} ||\bm{\nu}|| \delta_z(\phi, \bm{\gamma}^{\tilde{K}}) +  \delta_{\epsilon}(\bm{\gamma}^{\tilde{K}}) + \sqrt{\tilde{K}} \delta_f(\bm{\gamma}^{\tilde{K}}, \bm{\nu}).
\end{equation}\\
\end{proposition}
%\noindent \textit{Proof}: See Appendix \ref{appendix: propNA_lbd}. $\square$

The term $||\bm{\nu}|| \delta_z(\phi, \bm{\gamma}^{\tilde{K}})$ corresponds to Consequences (i), and $\delta_{\epsilon}(\bm{\gamma}^{\tilde{K}}) + \sqrt{\tilde{K}} \delta_f(\bm{\gamma}^{\tilde{K}}, \bm{\nu})$, which serves as a necessary condition to achieve a $(\delta_f, \delta_{\epsilon})$-generalizable SCM with small $\delta_f$ and $\delta_{\epsilon}$, corresponds to Consequences (ii).

\section{Causal Discovery}
\label{section: Causal Discovery}

\subsection{Discovering $(\delta_f, \delta_{\epsilon})$-Generalizable Causal Features}
\label{section: unravelInv}
The causal discovery corresponds to solving a constrained optimization problem:
\begin{equation}
\label{optimization problem by objs}
    \begin{split}
        &\min_{\phi, \bm{\nu}} \mathcal{L}_{\text{pred}}(\bm{\nu}, \phi)\\
        \text{s.t. \hspace{1em}} &\begin{cases}
            \mathcal{L}_{\text{res}}(\phi) = \xi,\\
            \mathbb{E}_{t_{-i}}[\phi(X)] = \mathbf{0} \hspace{1em} \forall i=1,\cdots, T,\\
            \mathbb{E}_{t_{-i}}[\phi(X)\phi(X)^{\top}] = \mathbf{I} \hspace{1em} \forall i=1,\cdots, T.
        \end{cases}
    \end{split}
\end{equation}
Here, $\mathcal{L}_{\text{pred}}(\bm{\nu}, \phi) := \frac{1}{T}\sum_{i=1}^T \mathbb{E}_{t_{-i}}[(R - \phi(X)^{\top} \bm{\nu})^2]$ represents the average mean squared error (MSE) over source domains. Minimizing $\mathcal{L}_{\text{pred}}(\bm{\nu}, \phi)$ aids in reducing the term $\frac{1}{T}\sum_{i=1}^T \textbf{err}_{t_{-i}}(g)$ in (\ref{errorbound}). Additionally, $\mathcal{L}_{\text{res}}(\phi) := \frac{1}{T}\sum_{i=1}^T \mathbb{E}_{t_{-i}}[(R - \phi(X)^{\top} \bm{\nu}^{\phi}_{t_{-i}})^2]$ is a measure of the residual size, where $\bm{\nu}^{\phi}_{t_{-i}} := \mathbb{E}_{t_{-i}}[\phi(X) \phi(X)^{\top}]^{-1} \mathbb{E}_{t_{-i}}[\phi(X) R]$ is the analytic solution of $\min_{\bm{\nu}' \in \mathbb{R}^{\tilde{K}}} \mathbb{E}_{t_{-i}}[(R - \phi(X)^{\top} \bm{\nu}')^2]$. We denote $\xi > 0$ as the residual level, which constrains the average variance of residuals when discovering causal features. Moreover, the first and second moments of $\phi(X)$ are constrained to ensure zero means, unit variances, and zero covariances for the entries of $\phi(X)$. This alignment not only matches the first two moments of $\phi(X)$ with the causal features $\tilde{\mathbf{Z}}$ of the transformed SCM (\ref{subportion}), but also helps approximately align the marginal distributions of $\phi(X)$ across all domains to meet the condition $\mathbb{P}_{t_0}(\phi(x)) = \mathbb{P}_{t_{-i}}(\phi(X))$ in Proposition \ref{prop:wass}.\footnote{This is because the first two moments of $\mathbb{P}_{t_{-i}}(\phi(x))$ are identical for $i=0, 1,\cdots, T.$} Notably, the constraints in (\ref{optimization problem by objs}) define a function space for $\phi$ that is independent of $\bm{\nu}$.

We explore under which conditions (\ref{optimization problem by objs}) is able to properly discover the causal features. We provide a rigorous analysis based on the comparison between causal features and \textit{spurious features}. That is, our SCM may contain \textit{spurious invariance}, which can misguide $\phi(X)$ towards identifying the features resulting from the non-causal factor $X^a$ instead of the true causal features $\tilde{\mathbf{Z}}$. Intuitively, the greater the variation in the relationship between \(X^a\) and \(R\) across different domains, the less significant is the spurious invariance in the SCM. This idea is formalized in the \textit{Causal/Non-Causal Deviation} Definition:\\

\begin{definition}
\label{def1}
(\textit{Causal/Non-Causal Deviation}.) Given a residual level $\xi>0$ and a dimensionality $\tilde{K}\geq1$, we denote $\Omega^{\phi}_{\xi, \tilde{K}}$ as a family of $\phi$ satisfying the constraints of (\ref{optimization problem by objs}). Let $\Psi_c, \Psi_{nc} \subset \Omega^{\phi}_{\xi, \tilde{K}}$ where $\Psi_c$ is the subset of $\phi$ that discover causal features (i.e., $\phi(X) = \tilde{\mathbf{Z}}$ for $\tilde{\mathbf{Z}}$ in the sense of Section \ref{section: Transformed_SCMs}) and $\Psi_{nc} := \Omega^{\phi}_{\xi, \tilde{K}} \backslash \Psi_c$. If $\Psi_c$ is non-empty, then the Causal and Non-Causal Deviations are defined, respectively, as:
\begin{equation}
\begin{split}
\label{def:C}
    \Delta_{c} &:= \min_{\phi \in \Psi_c, \bm{\nu} \in \mathbb{R}^{\tilde{K}}} \frac{1}{T}\sum_{i=1}^T ||\bm{\nu}^{\phi}_{t_{-i}} - \bm{\nu}||^2,\\
    \Delta_{nc} &:= \min_{\phi \in \Psi_{nc}, \bm{\nu} \in \mathbb{R}^{\tilde{K}}} \frac{1}{T}\sum_{i=1}^T ||\bm{\nu}^{\phi}_{t_{-i}} - \bm{\nu}||^2.
\end{split}
\end{equation}\\
\end{definition}

$\Delta_c$ and $\Delta_{nc}$ quantify the variance of coefficients of causal and spurious features, respectively, across source domains. Given this definition, Proposition \ref{prop:Lpred} elucidates that $\Delta_c < \Delta_{nc}$ serves as a sufficient condition for our causal discovery method.\\

\begin{proposition}
\label{prop:Lpred}
     Given $\Omega^{\phi}_{\xi,\tilde{K}}$ for $\xi>0$ and $\tilde{K} \geq 1$, suppose $\Psi_c\subset \Omega^{\phi}_{\xi,\tilde{K}}$ is non-empty, and the SCM (\ref{SCM}) is $(\delta_f, \delta_{\epsilon})$-generalizable for $\delta_f \geq \sqrt{\Delta_c}$ and $\delta_{\epsilon} \geq \sqrt{\xi}$. If $\Delta_c < \Delta_{nc}$, then the minimal solution of (\ref{optimization problem by objs}) discovers $(\delta_f, \delta_{\epsilon})$-generalizable causal features.\\
\end{proposition}
%\noindent \textit{Proof}: See Appendix \ref{appendix: propLpred}. $\square$

Both $\Delta_c$ and $\Delta_{nc}$ depend on the selection of the training horizon as well as how the source domains are constructed. However, it is non-trivial to estimate their values or find the best collection of domains. We leave this as an open question, and in this work, we assume that $\Delta_{nc}$ is low considering the large number of domains in our settings.

\subsection{Dimensionality of Discovered Causal Features} 
We delve deeper into the role of the dimensionality $\tilde{K}$ of $\phi(x)$. It is argued that in the IRM framework \cite{arjovsky2019invariant}, the higher-dimensional case has no representation advantage over the 1-dimensional counterpart. Therefore, their proposed algorithm only focuses on scalar predictors. This is because a $\tilde{K}$-dimensional model $\phi$ for $\tilde{K} > 1$ can be restructured into a 1-dimensional $\tilde{\phi}$, as described by:
\begin{equation}
\label{dimTrans}
\tilde{\phi}(x) \tilde{\nu} = \underbrace{\left(\frac{\phi(x)^{\top} \bm{\nu}}{||\bm{\nu}||}\right)}_{\tilde{\phi}(x)} \underbrace{||\bm{\nu}||}_{\tilde{\nu}}.
\end{equation}
However, this is not our case due to the constraints of (\ref{optimization problem by objs}), and we do gain benefits by increasing the dimensionality $\tilde{K}$. Proposition \ref{prop:dim} demonstrates that a lower objective value with a lower residual level is attainable by increasing the output dimensionality of $\phi$.\\

\begin{proposition}
\label{prop:dim}
For $1 \leq \tilde{K}_1 < \tilde{K}_2 \leq K$, let $\Psi^c_{\tilde{K}} := \{\phi: \mathbb{X} \rightarrow \mathbb{R}^{\tilde{K}} \mid \phi \text{ discovers causal features} \}$. Suppose $\Psi^{c}_{\tilde{K}_1} \subset \{\mathbf{L} \circ \phi \mid \phi \in \Psi^c_{\tilde{K}_2}, \mathbf{L} \in \mathbb{R}^{\tilde{K}_1 \times \tilde{K}_2}\}$. For any $\xi > 0$ and any $\phi \in \Psi^{c}_{\tilde{K}_1}$ that satisfies the constraints of (\ref{optimization problem by objs}) induced by $\xi$, there exists $\xi' > 0$ and $\phi' \in \Psi^{c}_{\tilde{K}_2}$, which satisfies the constraints of (\ref{optimization problem by objs}) induced by $\xi'$, such that $\xi' < \xi$ and $\min_{\bm{\nu} \in \mathbb{R}^{\tilde{K}_2}} \mathcal{L}_{\text{pred}}(\bm{\nu}, \phi') \leq \min_{\bm{\nu} \in \mathbb{R}^{\tilde{K}_1}} \mathcal{L}_{\text{pred}}(\bm{\nu}, \phi)$ where the set of equality has measure zero.\\
\end{proposition}
%\noindent \textit{Proof}: See Appendix \ref{appendix: propdim}. $\square$

The necessary condition $\Psi^{c}_{\tilde{K}_1} \subset \{\mathbf{L} \circ \phi \mid \phi \in \Psi^c_{\tilde{K}_2}, \mathbf{L} \in \mathbb{R}^{\tilde{K}_1 \times \tilde{K}_2}\}$ assumes that the representation power of $\phi' \in \Psi^c_{\tilde{K}_2}$ is sufficient. Indeed, by increasing the dimensionality, it is more likely to violate that necessary condition. For instance, if the input $X$ only contains information on $\tilde{K}_1$ factors, then $\Psi_{\tilde{K}_2}^c$ for $\tilde{K}_2 > \tilde{K}_1$ will be an empty set. Therefore, it is not plausible to choose arbitrarily large $\tilde{K}$.

\subsection{Learning Objectives}
\label{section: learning}
We implement $\phi$ using neural networks and train it to fulfill the objectives of (\ref{optimization problem by objs}). Firstly, For some residual level $\xi>0$, it is easy to verify that $\mathcal{L}_{res}(\phi) = \xi$ is equivalent to $\mathcal{L}_{res}'(\phi) = -\xi'$ for $\xi' := 1-\xi$ if $\phi(X)$ is domain-wisely standardized, where
\begin{equation}
    \label{Lexp2}
    \mathcal{L}_{res}'(\phi) := -\frac{1}{T}\sum_{i=1}^T ||\bm{\nu}^{\phi}_{t_{-i}}||^2.
\end{equation}
Secondly, inspired by the proof of Proposition \ref{prop:Lpred}, when the constraints of (\ref{optimization problem by objs}) induced by $\xi$ are satisfied, $\mathcal{L}_{\text{pred}}(\bm{\nu}, \phi)$ is minimized if
\begin{equation}
    ||\bm{\nu}||^2 = \xi' \hspace{1em}\wedge\hspace{1em} \bm{\nu}^{\phi}_{t_{-i}} = \bm{\nu}, \quad \forall i=1,\dots, T.
\end{equation}
As a result, we parameterize $\bm{\nu}$ as $\bm{\nu}(\bm{l}) := \left( \xi' \cdot \textit{softmax}(\bm{l}) \right)^{\frac{1}{2}}$, where the logits $\bm{l}$ are learnable parameters initialized to zero. The corresponding objective function is defined as:
\begin{equation}
    \mathcal{L}_{\text{inv}}(\bm{l}, \phi) := \frac{1}{T} \sum_{i=1}^{T} ||\bm{\nu}_{t_{-i}}^{\phi} - \bm{\nu}(\bm{l})||^2.
\end{equation}

Further, to align the first two moments, we standardize all entries of $\phi(X_{t_{-i}})$ within each domain and introduce a regularization term inspired by  \cite{lin2021deep}:
\begin{equation}
\label{Ldist}
    \mathcal{L}_{alig}(\phi) = \frac{1}{T}\sum_{i=1}^T \text{tr}\left((\mathbb{E}_{t_{-i}}[\phi(X) \phi(X)^{\top}])^{-1}\right),
\end{equation}
where $\text{tr}(\cdot)$ denotes the trace of a matrix. Given that the entries of $\phi(X)$ are standardized, $\bm{\Sigma} := \mathbb{E}_{t_{-i}}[\phi(X) \phi(X)^{\top}]$ becomes a correlation matrix. Hence, $\text{tr}(\bm{\Sigma}^{-1}) = \text{tr}(\bm{\Lambda}^{-1})$ and $\text{tr}(\bm{\Lambda}) = \text{tr}(\bm{\Sigma}) = \tilde{K}$, where $\bm{\Lambda}$ is the diagonal matrix obtained by diagonalizing $\bm{\Sigma}$. Consequently, the optimization problem $\min_{\bm{\Sigma}} \text{tr}(\bm{\Sigma}^{-1})$ can be reformulated as $\min_{\bm{\Lambda}} \text{tr}(\bm{\Lambda}^{-1})$ under the constraint $\text{tr}(\bm{\Lambda}) = \tilde{K}$. The identity matrix is the unique minimal solution, making it the minimal solution for the original problem as well.

In addition, selecting a suitable value for $\xi'$ that achieves a balance between invariance and residual level can be challenging. To address this, we automate the selection of $\xi'$ by setting it equal to $-\mathcal{L}'_{\text{res}}(\phi)$ during the training process. Specifically, the overall learning objective becomes:
\begin{equation}
\label{Lobj}
\begin{split}
    \min_{\phi, \bm{l}} \mathcal{L}'_{\text{res}}(\phi) &+ \lambda_1 \mathcal{L}_{\text{inv}}(\bm{l}, \phi) + \lambda_2 \mathcal{L}_{\text{alig}}(\phi),\\
    \text{subject to } \xi' &= -\mathcal{L}'_{\text{res}}(\phi),
\end{split}
\end{equation}
where $\lambda_1, \lambda_2 > 0$ are hyperparameters. Intuitively, $\lambda_1$ controls the strength of invariance, with a larger $\lambda_1$ leading to a smaller value for $\xi'$.

Recall that $\bm{\nu}^{\phi}$ is estimated using all samples within a domain, and $\xi'$ is set to the average of $||\bm{\nu}^{\phi}_{t_{-i}}||^2$ across all domains. However, due to GPU memory limitations, it is impractical to include all $T$ domains in a single batch during training, requiring the computation of gradients in mini-batches. To address this, we include $T_b$ domains and $N_b$ samples from each domain within a batch, where $1 \leq T_b \ll T$. The value of $\xi'$ is then updated using a momentum-based approach:
\begin{equation}
\label{eqn: MomentumUpdate}
\xi'_{j + 1} = \alpha \xi'_{j} + (1 - \alpha) \hat{\xi'}_b,
\end{equation}
where $j$ denotes the iteration step, $\alpha \in (0, 1)$ controls the momentum, and $\xi'_b$ is the estimated average of $||\bm{\nu}^{\phi}_{t_{-i}}||^2$ within the batch.

\section{Experiments}
\label{section: experiments}
\textbf{Datasets.} We present a comprehensive analysis of the Chinese A-share market over the period from January 4, 2010, to March 4, 2024, encompassing more than 4000 stocks. Our study employs a daily model utilizing the most recent 20-day Open, High, Low, Close, and Volume (OHLCV) data, as well as the 20-day signals from the \textit{Alpha191} dataset, as the raw inputs for each stock. The \textit{Alpha191} dataset, provided by GUOTAI JUNAN SECURITIES, consists of 191 quantitative 1-dimensional signals tailored for the Chinese market, each representing a distinct formula based on various market data such as prices and volumes.\footnote{The dataset is uploaded to the IEEE DataPort with a DOI: (10.21227/anjb-bq76).} Consequently, the inputs for each stock are structured as a $20 \times 196$ tensor. To mitigate the extensive computational burden, our methodology includes dynamically selecting the top 1000 largest companies by market capitalization, ensuring no missing values, to constitute our daily cross-sectional stock pools. We assess the effect of nonstationarity over the period from January 4, 2010, to March 4, 2024, while the OOS period from September 1, 2022, to March 4, 2024, is used for backtesting the performance of portfolios constructed under different causal discovery settings. To ensure tradability, the labels are computed through open prices:
\begin{equation}
    r_d^n := \frac{open^n_{d+2} - open^n_{d+1}}{open^n_{d+1}},
\end{equation}
where $open_t^n$ denotes the open price of stock $n$ at day $d$.

\textbf{Architectures.} Note that the scope of this paper focuses on the feature representation level, allowing flexibility in the choice of neural network architectures. To ensure universality, we consider the most common and well-studied architectures:
\begin{itemize}
    \item \textit{LSTM:} We employ a three-layer model where the first two layers are Bidirectional LSTMs \cite{schuster1997bidirectional} with 128 and 64 output neurons, respectively, each returning sequences. The final layer is an LSTM \cite{hochreiter1997long} with 32 output neurons that returns only the last step of the sequence. All layers utilize tanh activation functions for their outputs.
    \item \textit{ResNet1D:} This refers to a Residual Network \cite{he2016deep} with residual blocks based on 1-dimensional convolutional neural networks \cite{lecun1998gradient} designed to capture temporal features. Specifically, we define the Conv1D block as a 1-dimensional convolutional layer with same padding, followed by a Batch Normalization layer and a ReLU activation layer. A residual block is defined by stacking a Conv1D block and adding the block's output to its input, forming the residual connection. The ResNet1D architecture is constructed as follows: we begin with a Conv1D block with a stride of 1, a kernel size of 1, and 32 filters, applied to the input layer. This is followed by four residual blocks. The first two residual blocks have a stride of 1, a kernel size of 2, and 32 filters. The third block performs downsampling with a residual block where the Conv1D block has a stride of 2, a kernel size of 1, and 64 filters. The fourth residual block is similar to the first two blocks but with 64 filters, followed by a 1-dimensional Global Average Pooling layer. Finally, a fully connected layer with 32 output neurons and a tanh activation function is added.
    \item \textit{TFM:} This refers to a Transformer-like encoder \cite{vaswani2017attention}. The inputs are firstly mapped into the subspace using a 1-dimensional convolutional layer with a stride of 1, a kernel size of 1, 50 filters, and a linear activation function. Positional encodings are then added to the outputs. Following this, three Transformer blocks are stacked. The output layer is an LSTM layer with 32 output neurons and a tanh activation function. Each Transformer block consists of two sequential components: (i) a multi-head attention layer with 4 heads, each head having a size of 64, and (ii) a 1-dimensional convolutional layer with a stride of 1, a kernel size of 1, 50 filters, and a ReLU activation function. Both components are followed by layer normalization and their outputs are connected to the previous layer in a residual manner.
\end{itemize}
All outputs of the neural networks are projected into the $\tilde{K}$-dimensional space by adding a dense layer with $\tilde{K}$ output neurons, using a linear activation function and without biases. These architectures are used to parameterize $\phi$ in the causal discovery scenarios, where $\tilde{K}$ is a hyperparameter associated with $\bm{\gamma}^{\tilde{K}}$ as defined in (\ref{eqn: gamma}). They are also used to parameterize $g(x)$ directly in the baselines with $\tilde{K}=1$.

\textbf{Baselines.} We consider baseline algorithms where the prediction models $g(x)$ are directly parameterized by the listed architectures with $\tilde{K}=1$ without causal discovery. Domain knowledge is disregarded by pooling all samples into one mixed dataset and shuffling it. The gradients are computed by minimizing the mean squared error (MSE) estimated within mini-batches.

\textbf{Portfolio Performances.} It is natural to construct portfolios that update positions on a daily basis and analyze their performance. These portfolios involve taking long positions in stocks and hedging with a $1/N$ portfolio constructed by equally investing in daily stock pools consisting of $N$ stocks. For the long positions, the weight of each stock \(n\) is defined as \(\frac{rank(g(x^n_d))}{\sum_{n=1}^{N}rank(g(x^n_d))}\), where $rank(g(x^n_d))$ denotes the rank of stock $n$'s output signal at day $d$. The performance of these portfolios is assessed using the following metrics:
\begin{itemize}
    \item \textit{CAGR.} The Cumulative Annual Growth Rate (CAGR) is calculated as \(CAGR := (CAP_{end} / CAP_{start})^{\frac{252}{L}} - 1\), where \(CAP_{start}\) and \(CAP_{end}\) represent the capital at the beginning and end of the period, respectively, and \(L\) denotes the trading length in days.
    \item \textit{SR.} The Sharpe Ratio (SR) is calculated as \(SR := \sqrt{252}\frac{\mu_{port}}{\sigma_{port}}\), where \(\mu_{port}\) is the average daily return and \(\sigma_{port}\) is the standard deviation of daily returns.
\end{itemize}

Portfolios are constructed to examine performances over the OOS period from September 1, 2022, to March 4, 2024, employing a rolling strategy. Specifically, the model update frequency is set at 120 trading days (i.e., $(t_{-1}, t_0]$ consists of 120 days), so that we have 3 updates for each algorithm. Optimization of models occurs on the most recent 350-day window, partitioned such that the initial 300 days constitute the training set, while the subsequent 50 days serve as the validation set. Each source domain is constructed through 5 consecutive days, so that the training set is partitioned into 60 domains, denoted by $\mathcal{T}_{tr}$, and the validation set is partitioned into 10 domains, denoted by $\mathcal{T}_{val}$.

For the causal discovery algorithms, we consider the cases $\tilde{K} = 1$, 2, 3, and 4. The training settings are identical for all models on all training sets: Inputs are feature-wise min-max scaled on the training sets; We choose $\lambda_1=5$ and $\lambda_2=1$ for the hyperparameters; The model is optimized using the Adam optimizer \cite{kingma2014adam} with the learning rate set at $0.0002$; Each batch consists of 3 domains and 1000 samples from each domain; Training is early-stopped upon the stagnation of validation MSE over five successive epochs. We train 9 models for each algorithm and report the averages and standard deviations in Table \ref{tab:performance_metrics}. The baselin algorithms follow the same setting except for $\lambda_1$ and $\lambda_2$.

We analyze the results reported in Table \ref{tab:performance_metrics}. Regardless of architectures or subperiods, the causal discovery algorithms (i.e., $\tilde{K}\geq 1$) outperform the baseline algorithms in terms of both SR and CAGR. Among the baselines, we found that TFM performs the worst across all three subperiods. However, causal discovery rescues TFM: Transformer-like architectures with causal discovery outperform their counterparts in two of the subperiods. On the other hand, TFM performs the worst among all causal discovery algorithms during the subperiod between March 4, 2023, and August 28, 2023, where ResNet1D significantly outperforms. It seems different market periods prefer different inductive biases equipped by neural network architectures, and no single architecture dominates all OOS domains. 

\begin{figure}[t]
    \centering
    \includegraphics[width=0.6\linewidth]{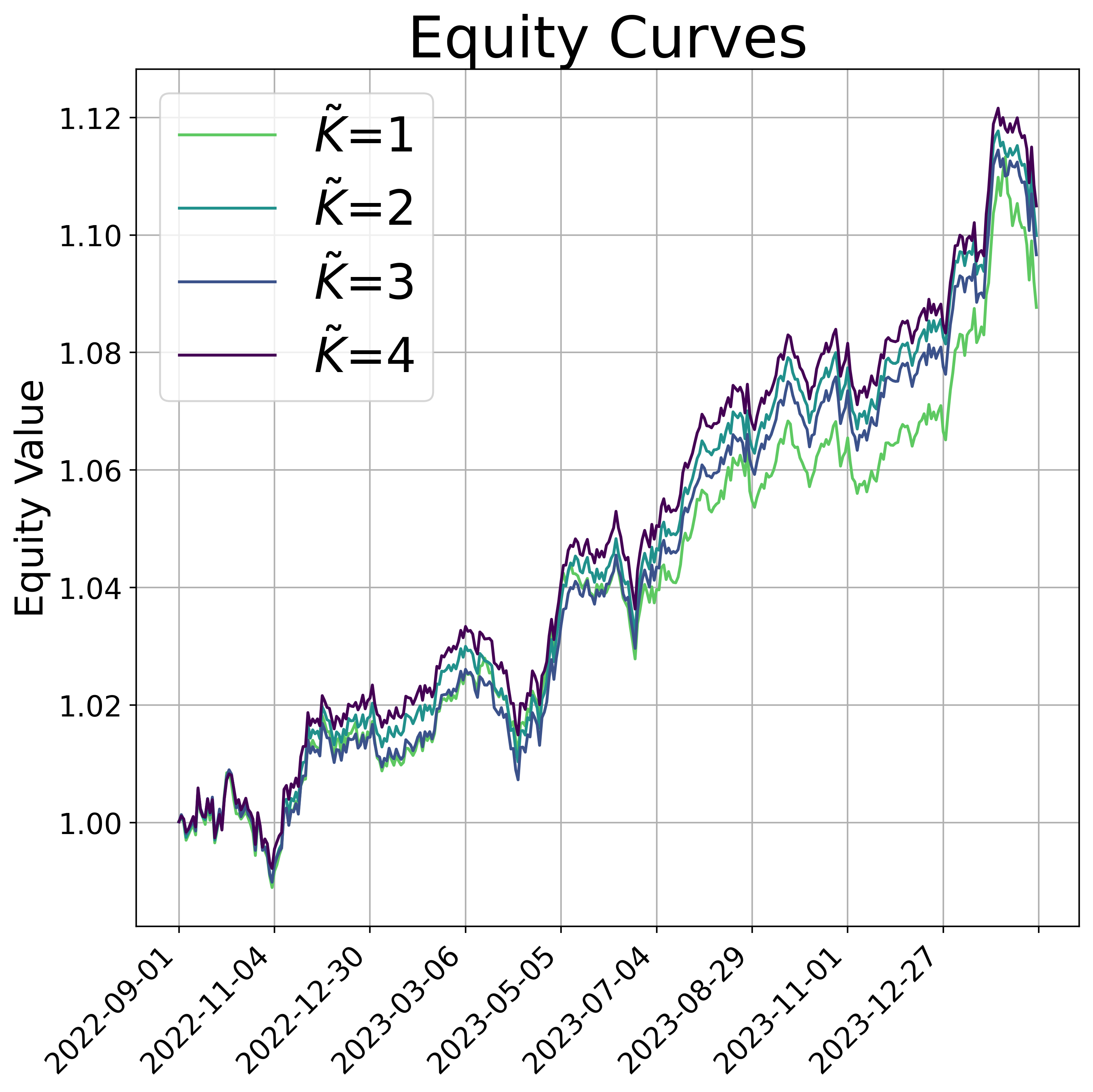}
    \caption{Equity curves of output ensembling over 9 runs $\times$ 3 architectures for $\tilde{K}=1$, 2, 3, and 4. The performance metrics are reported as follows: $\tilde{K}=1$: SR = 1.64, CAGR = 6.04\%; $\tilde{K}=2$: SR = 1.91, CAGR = 6.88\%; $\tilde{K}=3$: SR = 1.88, CAGR = 6.66\%; $\tilde{K}=4$: SR = 2.04, CAGR = 7.22\%.}
    \label{fig:ECs_dims}
\end{figure}

Additionally, the best results for each pair of architecture and subperiod occur at $\tilde{K}\geq 2$, with 6 out of 9 best performances occurring at $\tilde{K}=4$. This supports the idea that we can gain benefits by increasing $\tilde{K}$ in our causal discovery framework. To visualize, we ensemble the model outputs $g(x)$ over 9 runs $\times$ 3 architectures for each value of $\tilde{K}$ and construct portfolios according to the ensembled outputs. Figure \ref{fig:ECs_dims} shows the equity curves induced by the ensembled outputs, demonstrating that $\tilde{K}\geq 2$ significantly outperforms $\tilde{K}=1$, and $\tilde{K}=4$ results in the best SR and CAGR. 

\begin{table*}[!t]
\caption{Performance metrics over three subperiods. Metric values are reported as average values over 9 runs, with the best values bolded for each pair of architecture and subperiod. The standard deviations are reported in parentheses.}
\label{tab:performance_metrics}
\centering
\begin{tabular}{|l|cc|cc|cc|}
    \hline
    & \multicolumn{2}{c|}{2022.09.01-2023.03.03} & \multicolumn{2}{c|}{2023.03.04-2023.08.28} & \multicolumn{2}{c|}{2023.08.29-2024.03.04} \\
    & SR & CAGR & SR & CAGR & SR & CAGR \\
    \hline
    Baseline (ResNet1D) & 0.86 & 0.63\% & 0.86 & 0.01\% & 1.21 & 0.16\% \\
    & (0.81) & (0.59\%) & (1.04) & (1.06\%) & (0.33) & (0.37\%) \\
    ResNet1D ($\tilde{K}=1$) & 1.18 & 4.32\% & 2.24 & 7.47\% & 1.40 & 4.50\% \\
    & (0.12) & (0.67\%) & (0.10) & (0.49\%) & (0.51) & (1.41\%) \\
    ResNet1D ($\tilde{K}=2$) & 1.41 & 5.60\% & 2.40 & 8.30\% & 1.82 & 6.57\% \\
    & (0.19) & (0.38\%) & (0.12) & (0.52\%) & (0.10) & (0.42\%) \\
    ResNet1D ($\tilde{K}=3$) & 1.41 & 5.22\% & 2.50 & 8.37\% & 1.84 & 6.62\% \\
    & (0.15) & (0.69\%) & (0.18) & (0.57\%) & (0.09) & (0.49\%) \\
    ResNet1D ($\tilde{K}=4$) & \textbf{2.11} & \textbf{6.86\%} & \textbf{2.53} & \textbf{8.66\%} & \textbf{1.84} & \textbf{6.74\%} \\
    & (0.54) & (0.70\%) & (0.15) & (0.54\%) & (0.07) & (0.28\%) \\
    \hline
    Baseline (RNN) & 1.12 & 2.06\% & 1.34 & 3.52\% & 0.88 & 3.32\% \\
    & (1.31) & (2.07\%) & (1.27) & (2.43\%) & (1.03) & (2.88\%) \\
    RNN ($\tilde{K}=1$) & 1.38 & 6.21\% & 1.59 & 6.77\% & 0.96 & 4.09\% \\
    & (0.06) & (0.10\%) & (0.12) & (0.63\%) & (0.12) & (0.93\%) \\
    RNN ($\tilde{K}=2$) & 1.46 & 6.23\% & 1.86 & 7.63\% & 1.23 & \textbf{5.32\%} \\
    & (0.24) & (0.92\%) & (0.13) & (0.58\%) & (0.18) & (1.05\%) \\
    RNN ($\tilde{K}=3$) & \textbf{1.47} & \textbf{6.62\%} & 1.78 & 7.26\% & 1.33 & 4.89\% \\
    & (0.41) & (1.83\%) & (0.16) & (0.71\%) & (0.32) & (1.19\%) \\
    RNN ($\tilde{K}=4$) & 1.23 & 3.43\% & \textbf{2.07} & \textbf{8.00\%} & \textbf{1.38} & 5.22\% \\
    & (0.34) & (0.01\%) & (0.33) & (0.01\%) & (0.24) & (0.01\%) \\
    \hline
    Baseline (TFM) & 0.32 & 2.98\% & 0.01 & 0.28\% & 0.21 & 0.39\% \\
    & (0.23) & (2.40\%) & (0.94) & (2.30\%) & (0.91) & (2.33\%) \\
    TFM ($\tilde{K}=1$) & 1.66 & 7.57\% & 1.216 & 5.50\% & 1.976 & 8.00\% \\
    & (0.1) & (0.39\%) & (0.18) & (0.86\%) & (0.47) & (1.22\%) \\
    TFM ($\tilde{K}=2$) & 1.99 & \textbf{8.31\%} & 1.14 & 4.80\% & 2.17 & 7.94\% \\
    & (0.23) & (0.47\%) & (0.15) & (0.78\%) & (0.36) & (0.88\%) \\
    TFM ($\tilde{K}=3$) & \textbf{2.12} & 7.66\% & 1.34 & 5.53\% & 2.03 & 7.72\% \\
    & (0.23) & (0.47\%) & (0.428) & (1.65\%) & (0.403) & (0.98\%) \\
    TFM ($\tilde{K}=4$) & 1.91 & 7.12\% & \textbf{1.38} & \textbf{5.72\%} & \textbf{2.28} & \textbf{8.07\%} \\
    & (0.26) & (0.62\%) & (0.15) & (1.11\%) & (0.44) & (1.60\%) \\
    \hline
\end{tabular}
\end{table*}

\textbf{Evaluate Causal Discovery.} To directly evaluate our proposed Causal Discovery, we estimate the Wasserstein term from the error bound (\ref{errorbound}) between $\mathcal{T}_{tr}$ and $\mathcal{T}_{val}$. The validation domains serve as a surrogate for the domain $t_0$. Since the validation set is partitioned into 10 domains, the distance between each source domain $t_{-i}$ and the OOS domain $t_0$ is estimated by
\begin{equation}
    \begin{split}
        &\hat{\mathcal{W}}^1(\mathbb{P}_{t_0}(g(x), r), \mathbb{P}_{t_{-i}}(g(x), r))\\ 
        &:= \frac{1}{|\mathcal{T}_{val}|}\sum_{t \in \mathcal{T}_{val}} \hat{\mathcal{W}}^1(\mathbb{P}_{t}(g(x), r), \mathbb{P}_{t_{-i}}(g(x), r)).
    \end{split}
\end{equation}
The Wasserstein distance is estimated by solving a constrained linear programming problem \cite{villani2009optimal}, which is notoriously computationally complex (i.e., $\mathcal{O}(N^4)$). Therefore, the empirical Wasserstein distance between $t_{-i}$ and $t$ for $t \in \mathcal{T}_{val}$ is estimated batch-wise: We split both source and validation domains into 25 batches, each containing 200 samples; 25 batch-wise distances are computed and the total distance is obtained by taking the average.

\begin{figure}[!t]
    \centering
    \subfloat[]{
        \includegraphics[width=0.32\linewidth]{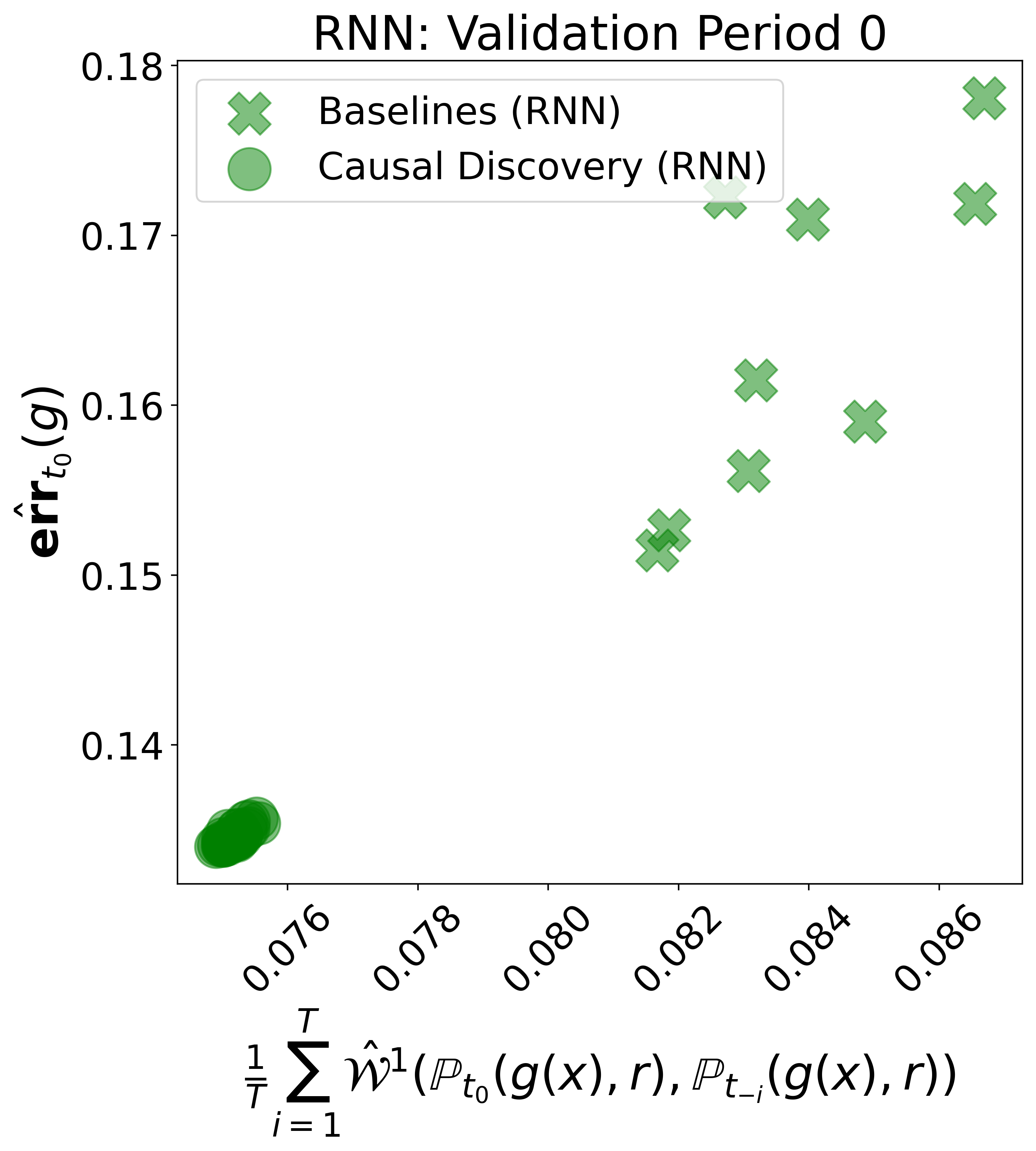}
    }
    \subfloat[]{
        \includegraphics[width=0.32\linewidth]{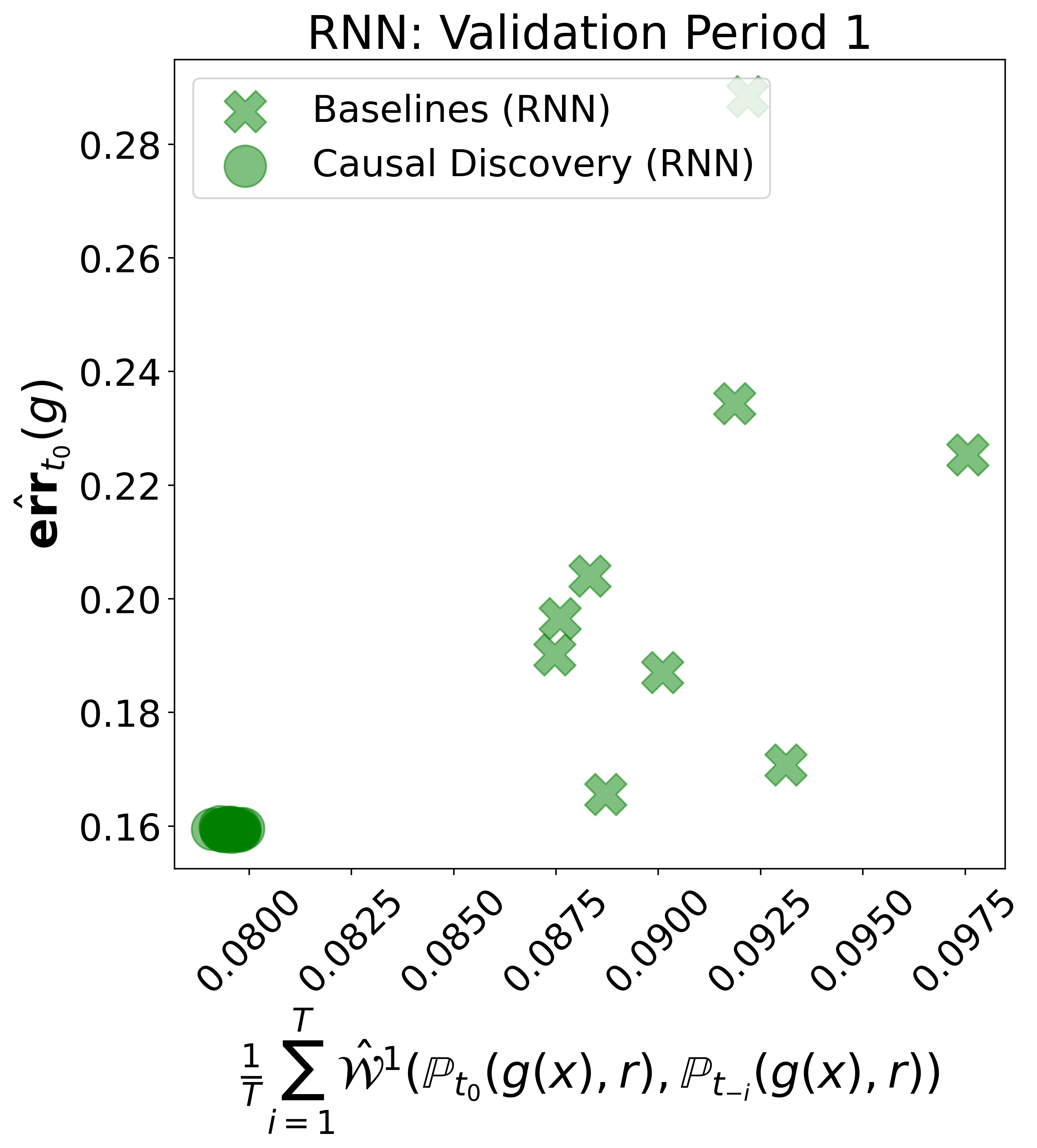}
    }
    \subfloat[]{
        \includegraphics[width=0.32\linewidth]{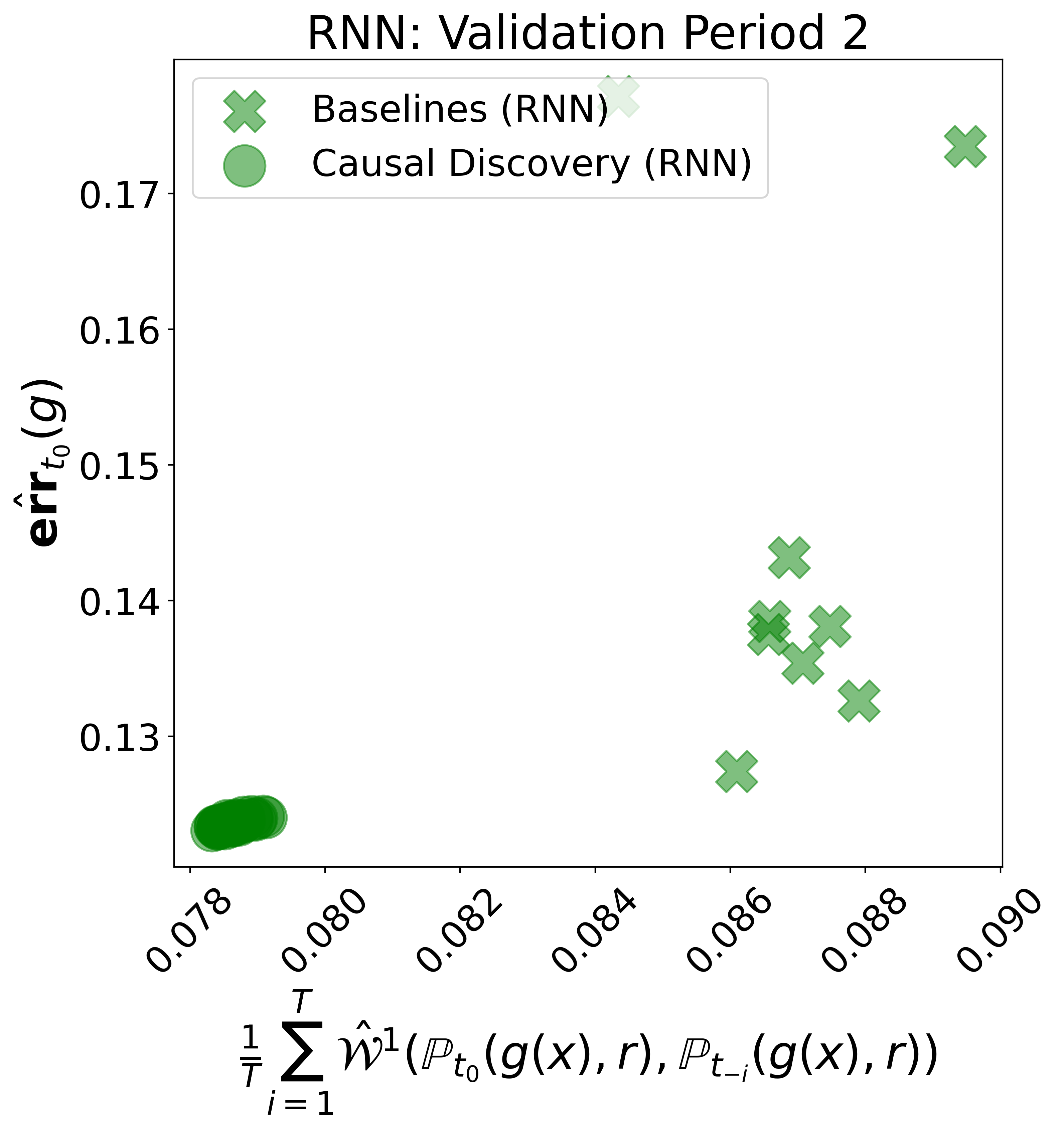}
    }\\

    \subfloat[]{
        \includegraphics[width=0.32\linewidth]{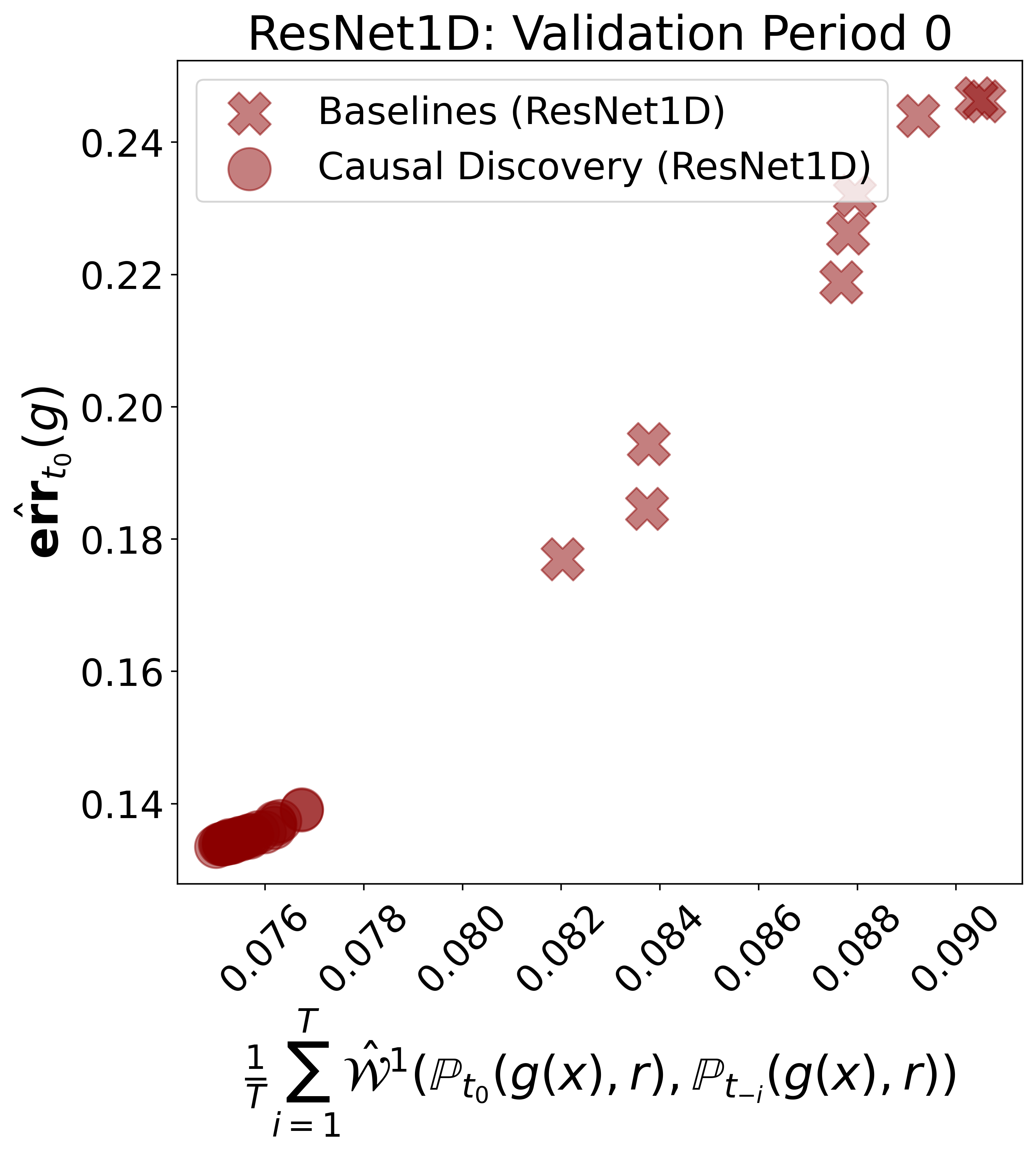}
    }
    \subfloat[]{
        \includegraphics[width=0.32\linewidth]{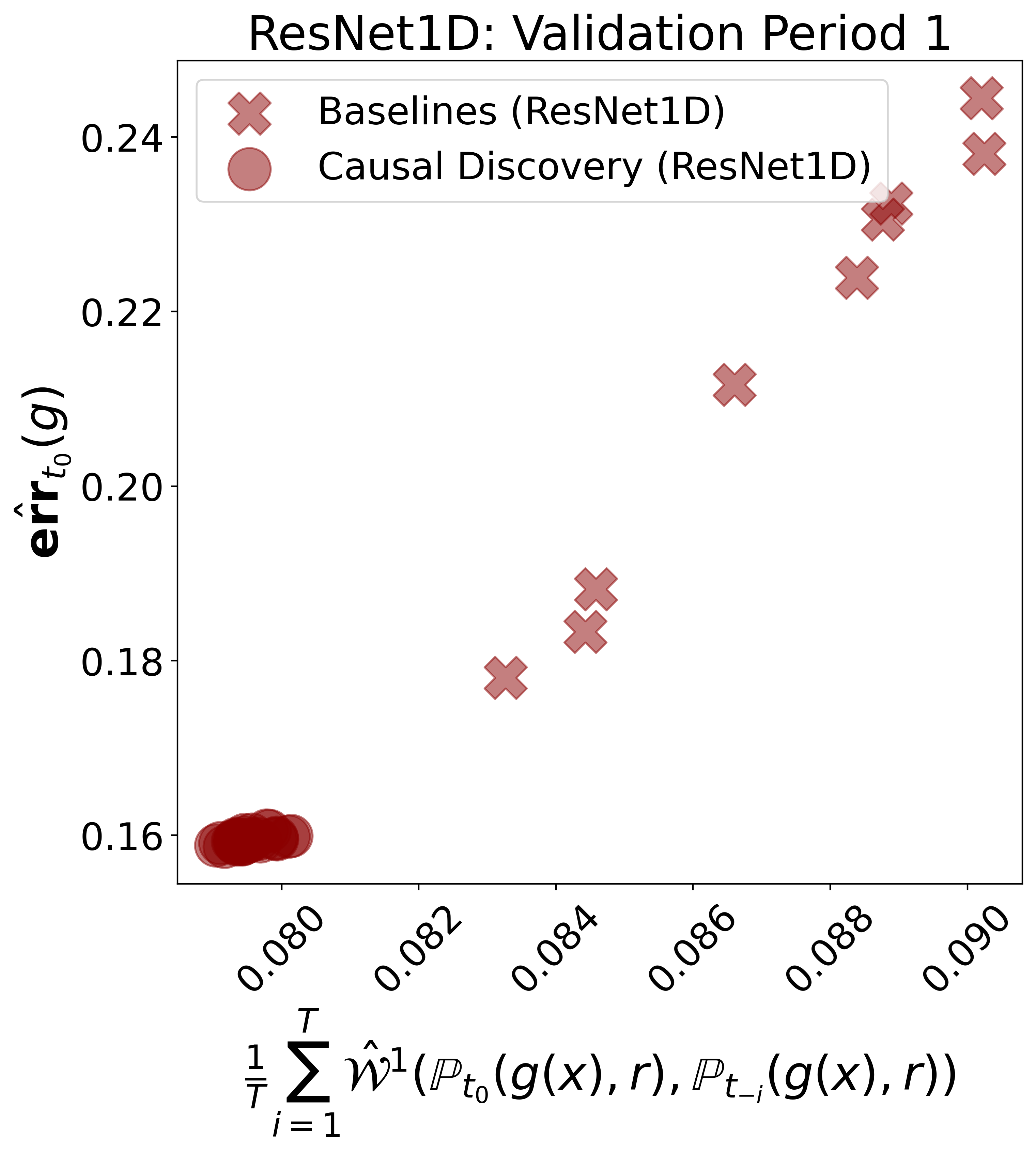}
    }
    \subfloat[]{
        \includegraphics[width=0.32\linewidth]{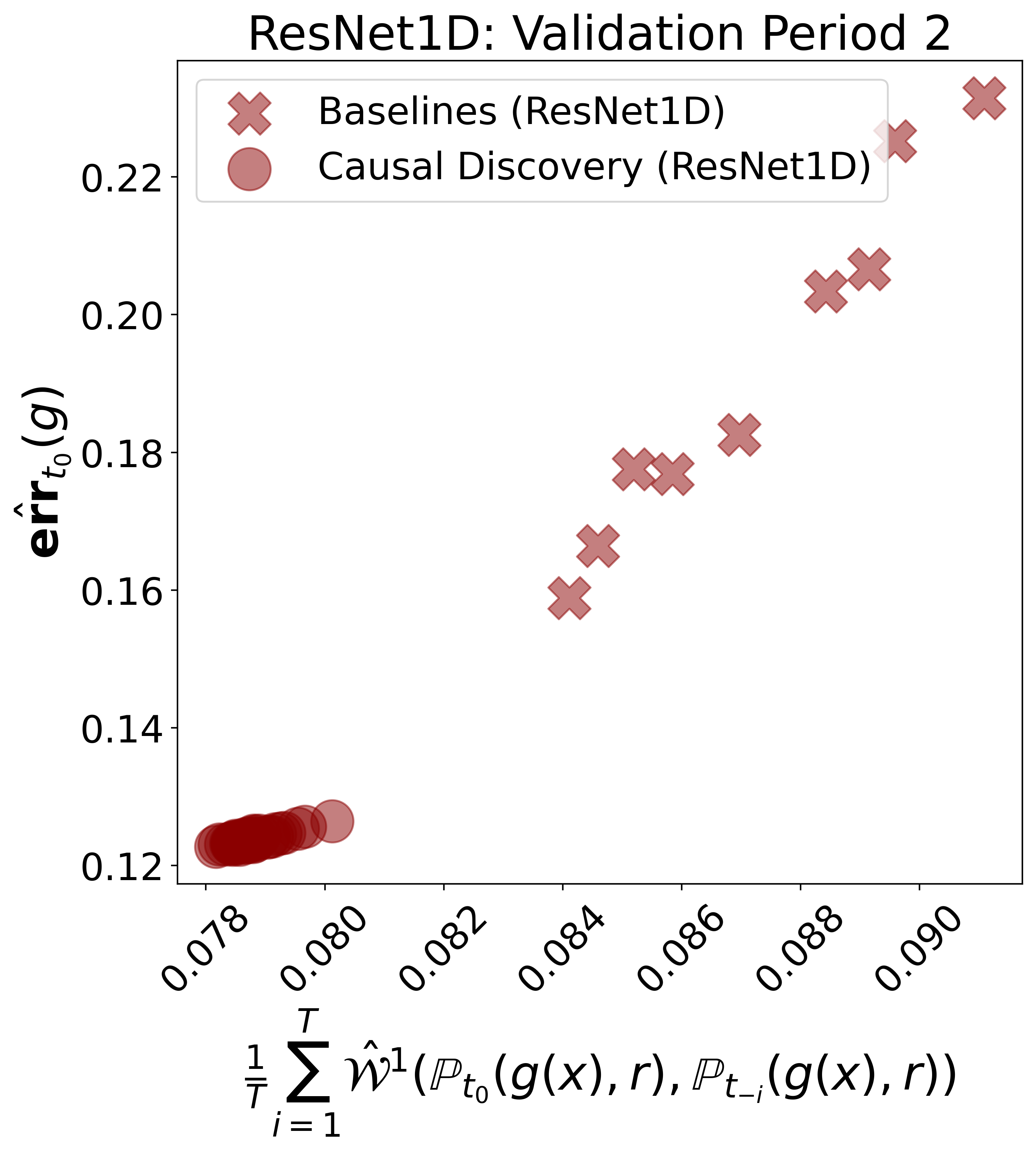}
    }
    \\

    \subfloat[]{
        \includegraphics[width=0.32\linewidth]{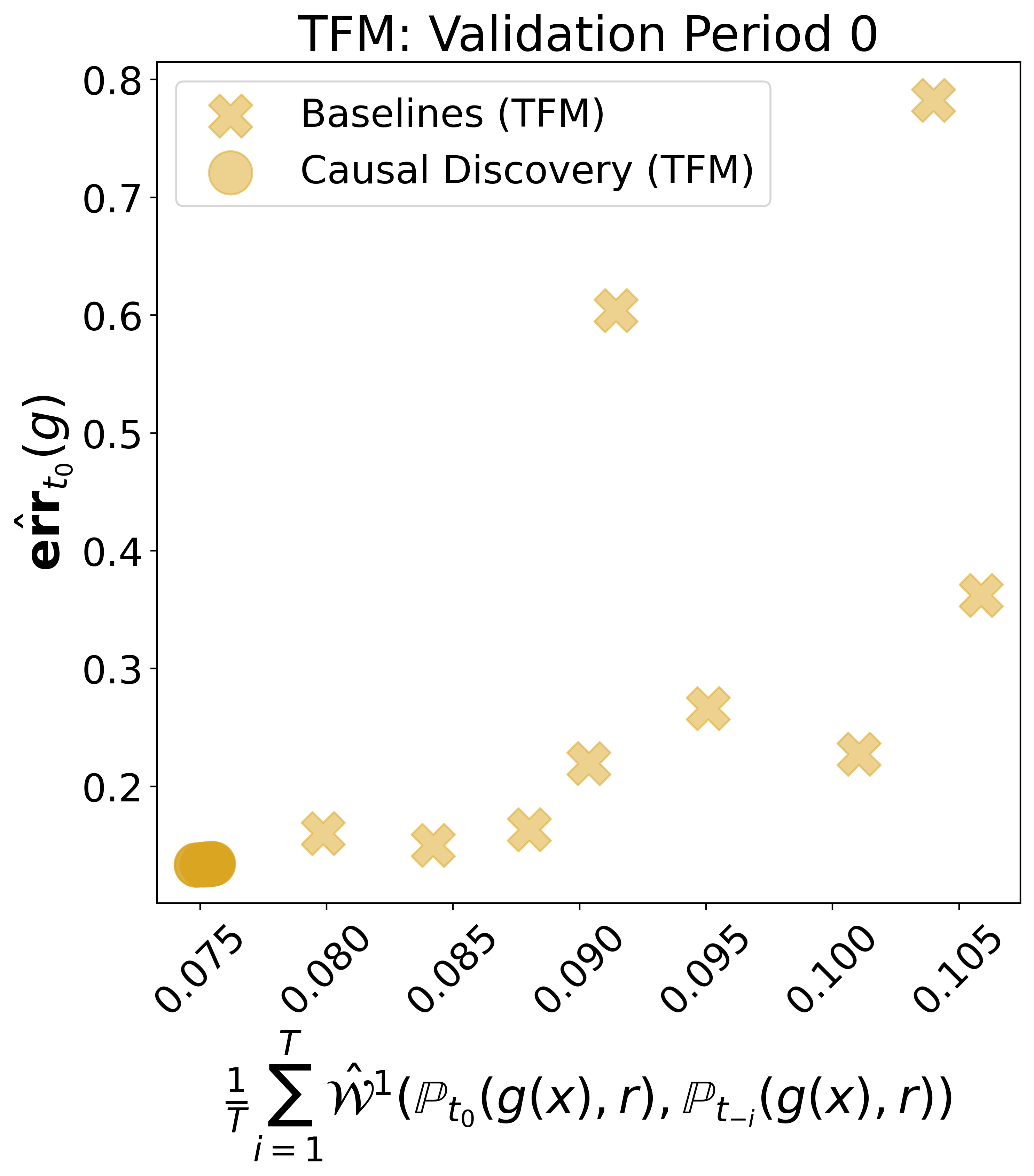}
    }
    \subfloat[]{
        \includegraphics[width=0.32\linewidth]{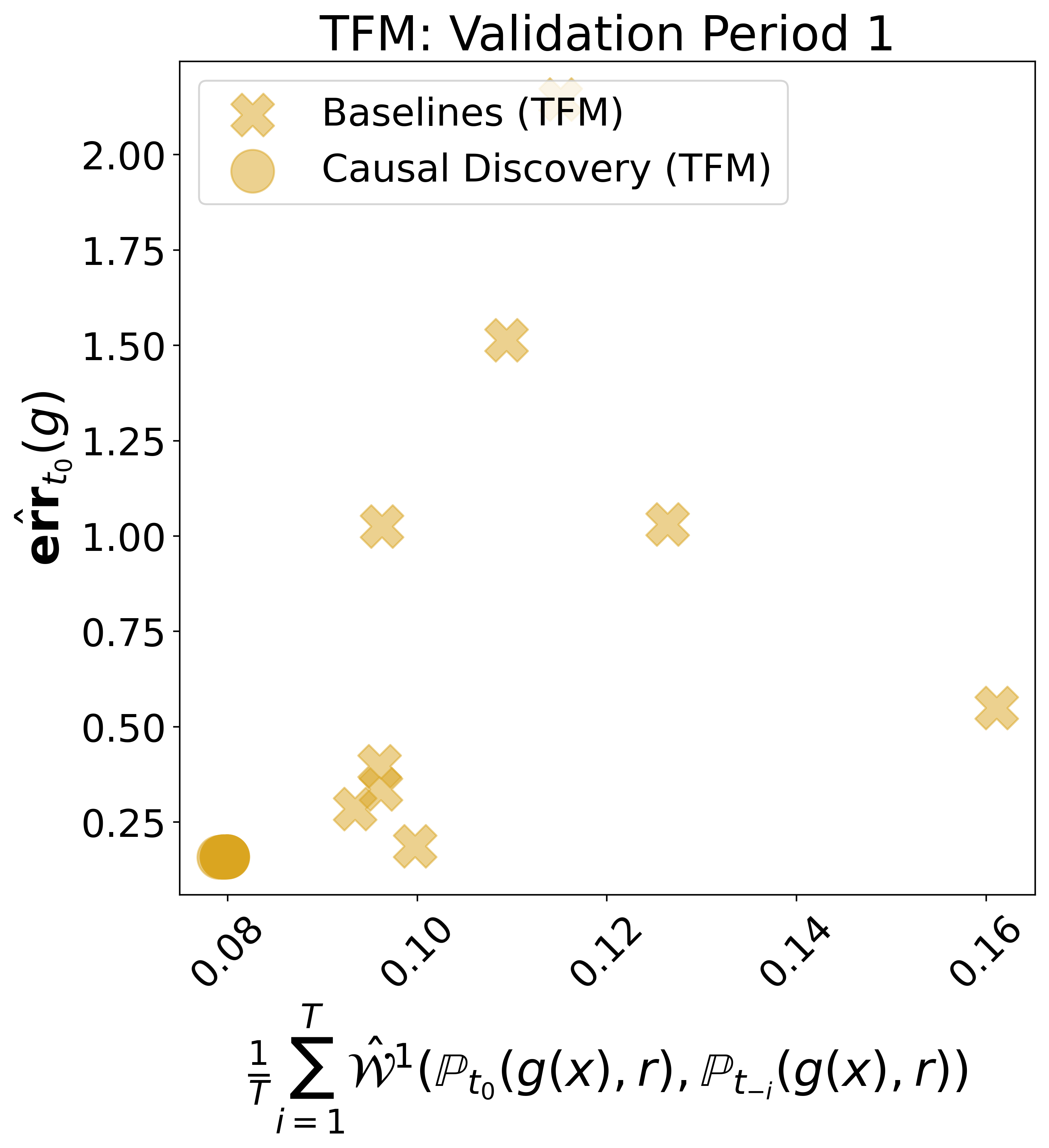}
    }
    \subfloat[]{
        \includegraphics[width=0.32\linewidth]{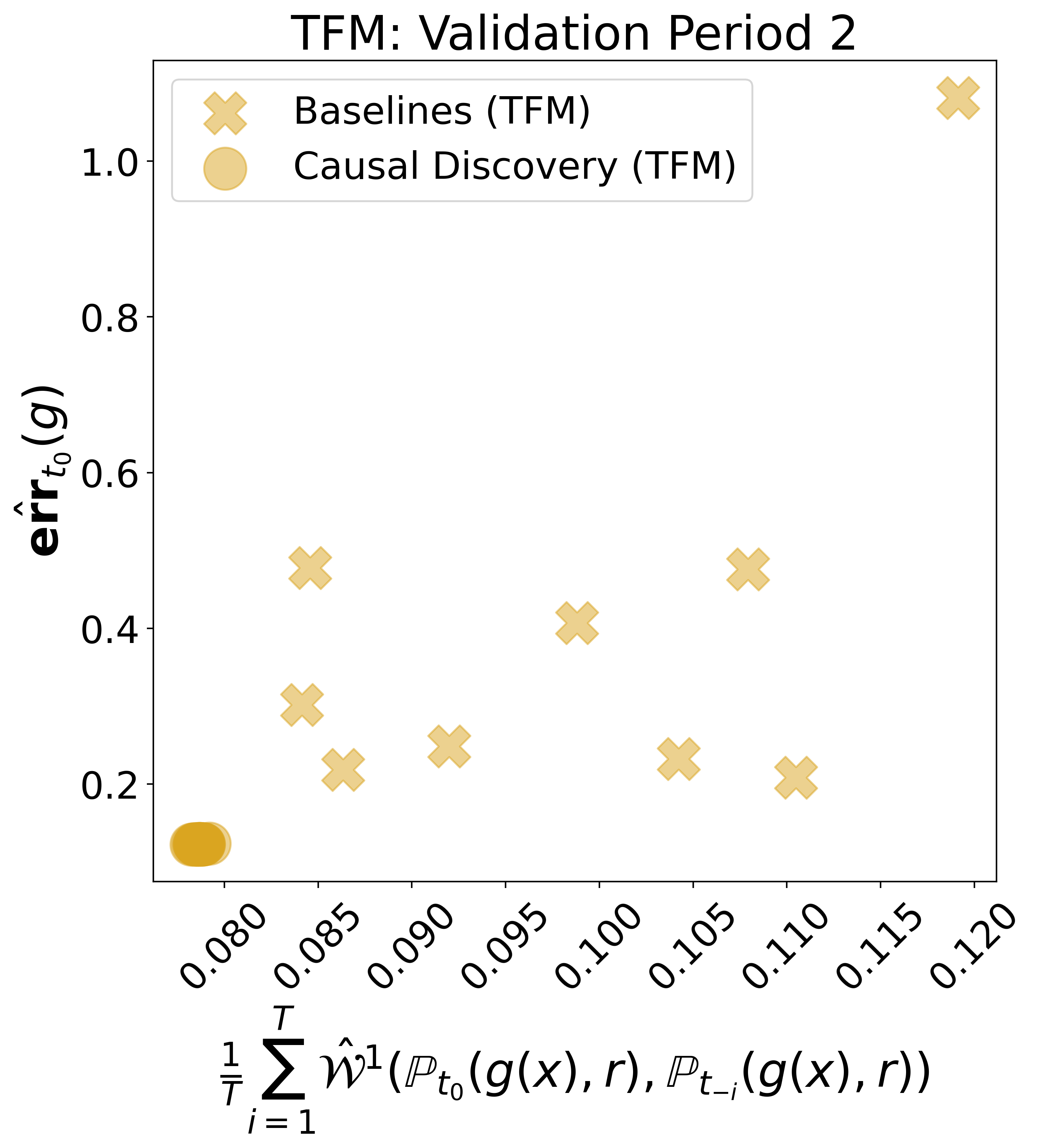}
    }

    \caption{OOS error versus Wasserstein term estimated on $\mathcal{T}_{tr}$ and $\mathcal{T}_{val}$. Subgraphs are plotted for 3 architectures (horizontal) $\times$ 3 validation periods (vertical). Within each subgraph, the baseline algorithms are marked with cross signs, while the causal discovery algorithms, involving the cases $\tilde{K}=1$, 2, 3, and 4, are marked with dots.}
    \label{fig:Wass}
\end{figure}

In Figure \ref{fig:Wass}, we plot the estimated Wasserstein term versus the OOS error, which is estimated by the average Mean Absolute Error (MAE) over validation domains. 9 cases are plotted (3 architectures $\times$ 3 validation periods), and each subplot consists of the baseline models, marked with cross signs, and the causal discovery models involving the cases $\tilde{K}=1$, 2, 3, and 4, marked with dots. A general pattern is that all the causal discovery models are clustered in the lower left corner, indicating lower OOS errors and lower values of the Wasserstein term, while the baseline models are scattered more randomly and far from the causal discovery clusters.

\begin{figure}[!t]
    \centering
    \subfloat[]{
        \includegraphics[width=0.32\linewidth]{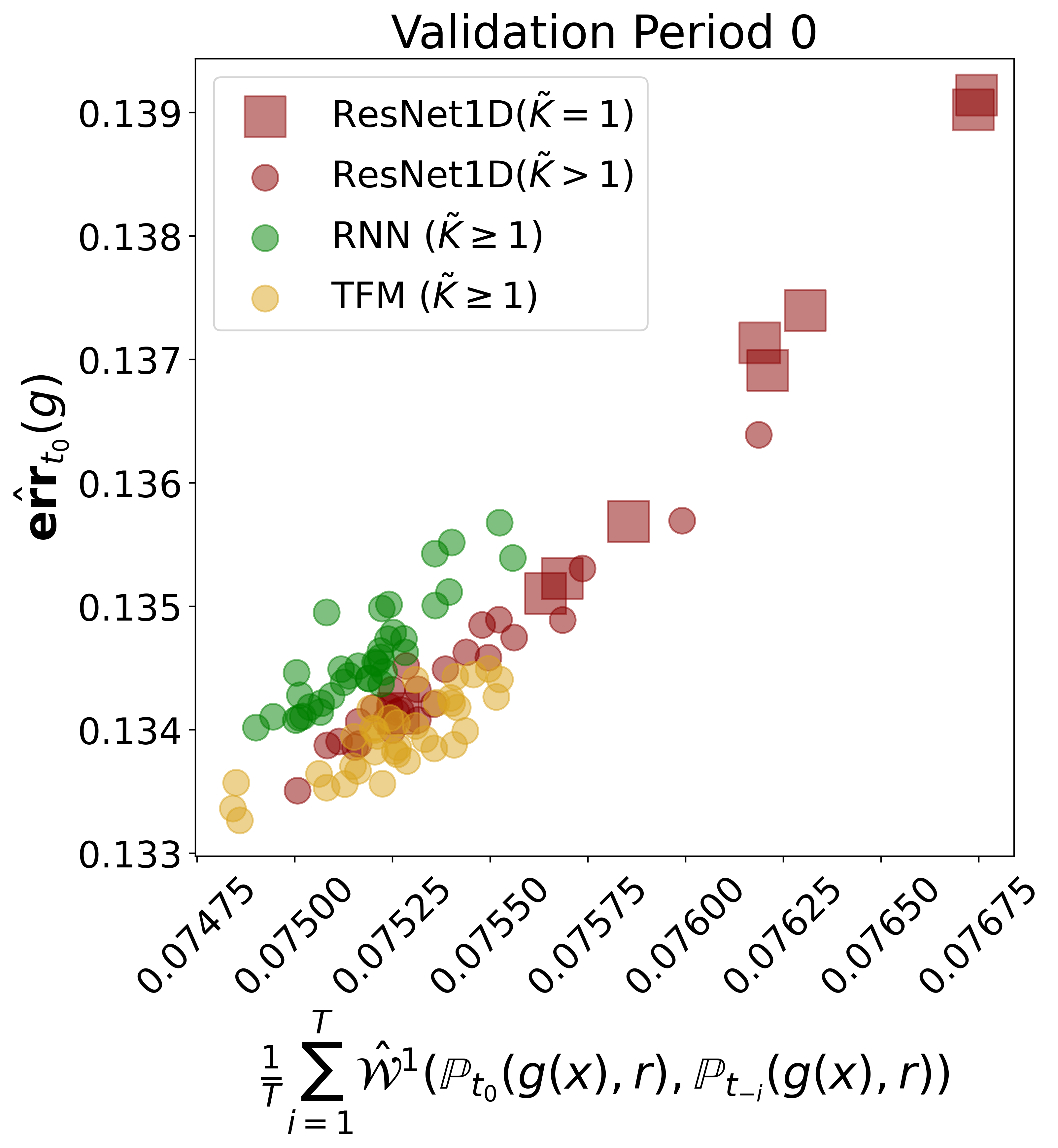}
        \includegraphics[width=0.32\linewidth]{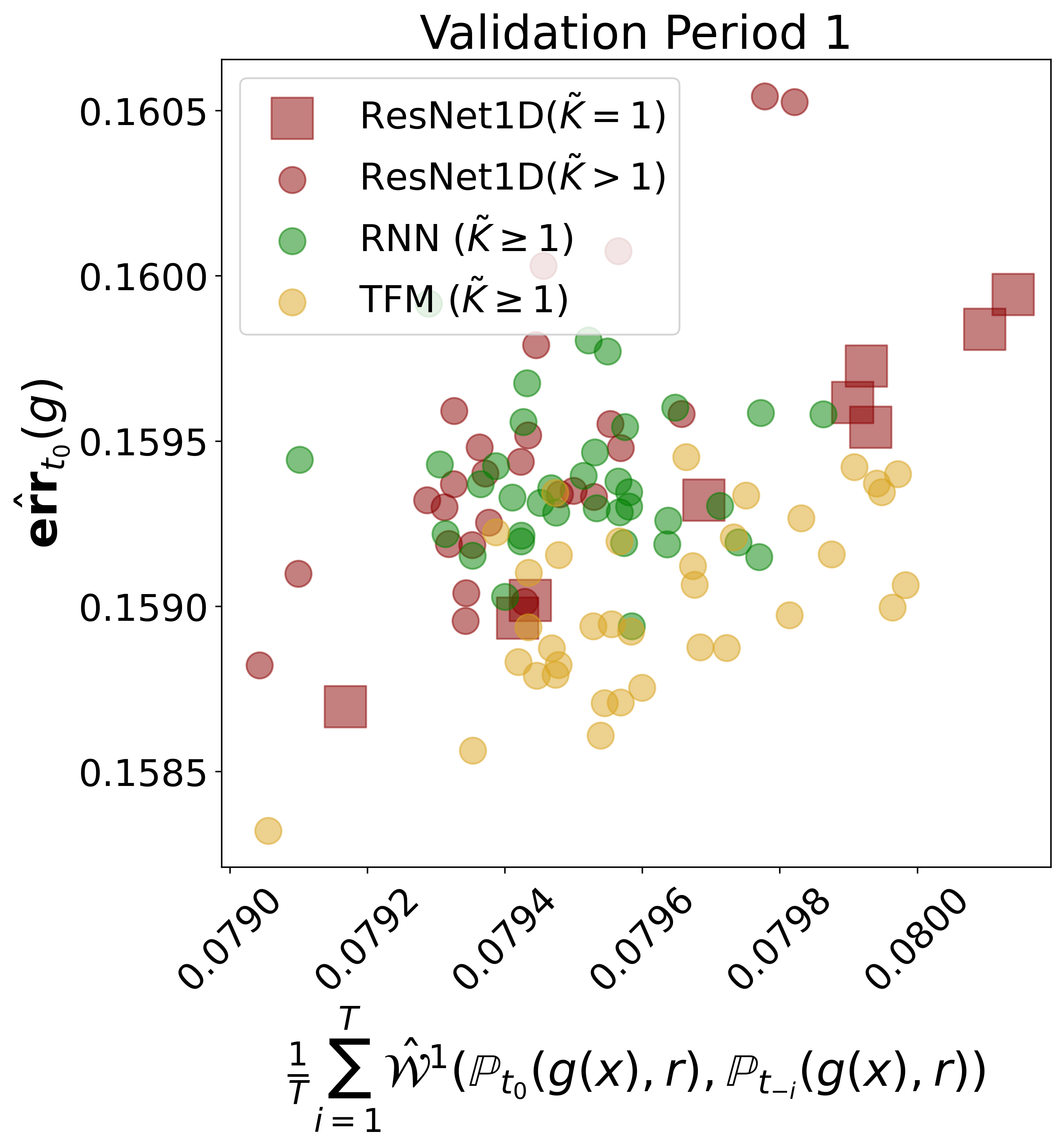}
        \includegraphics[width=0.32\linewidth]{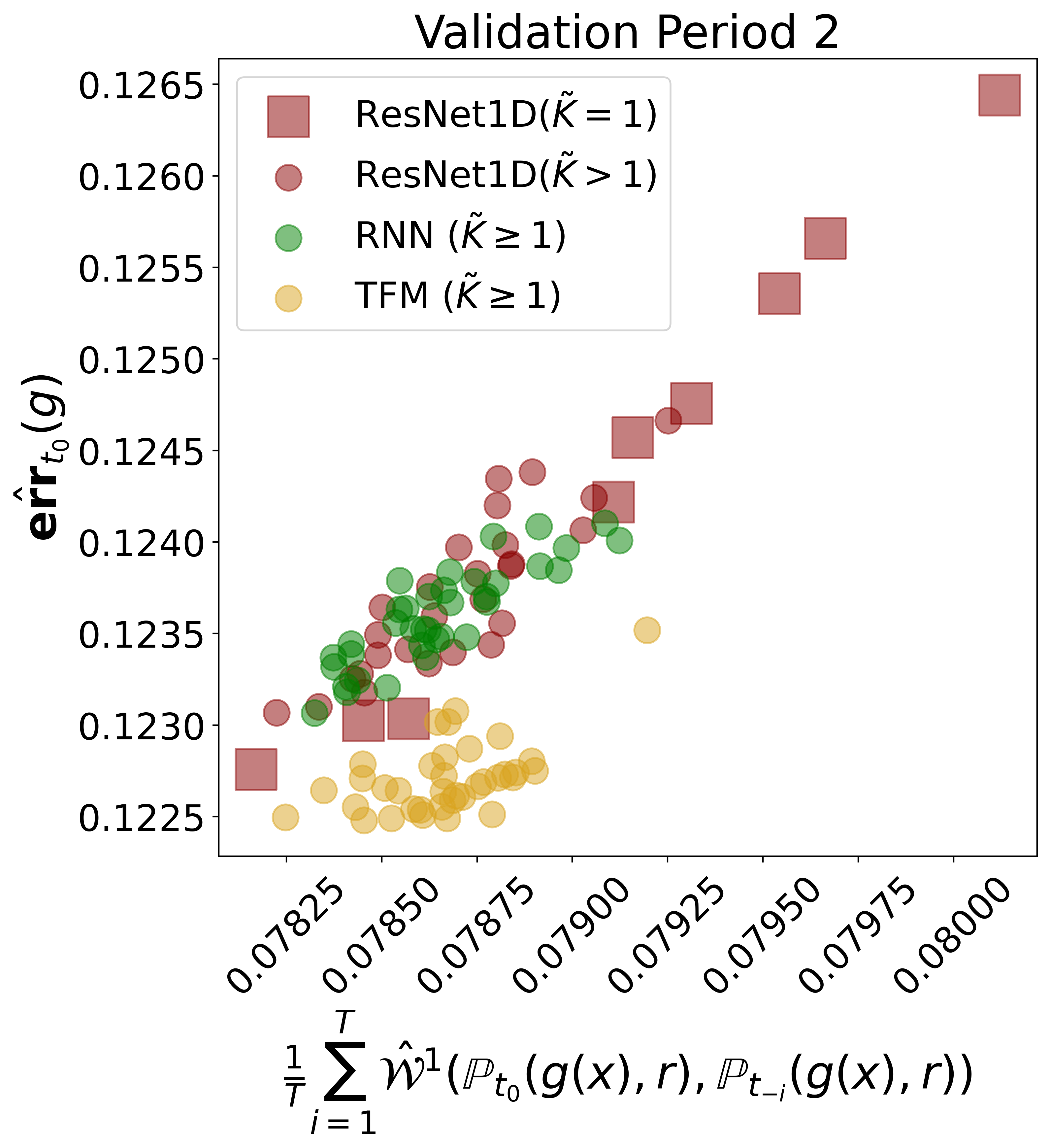}
        \label{fig:Wass_2a}
    }

    \subfloat[]{
        \includegraphics[width=0.32\linewidth]{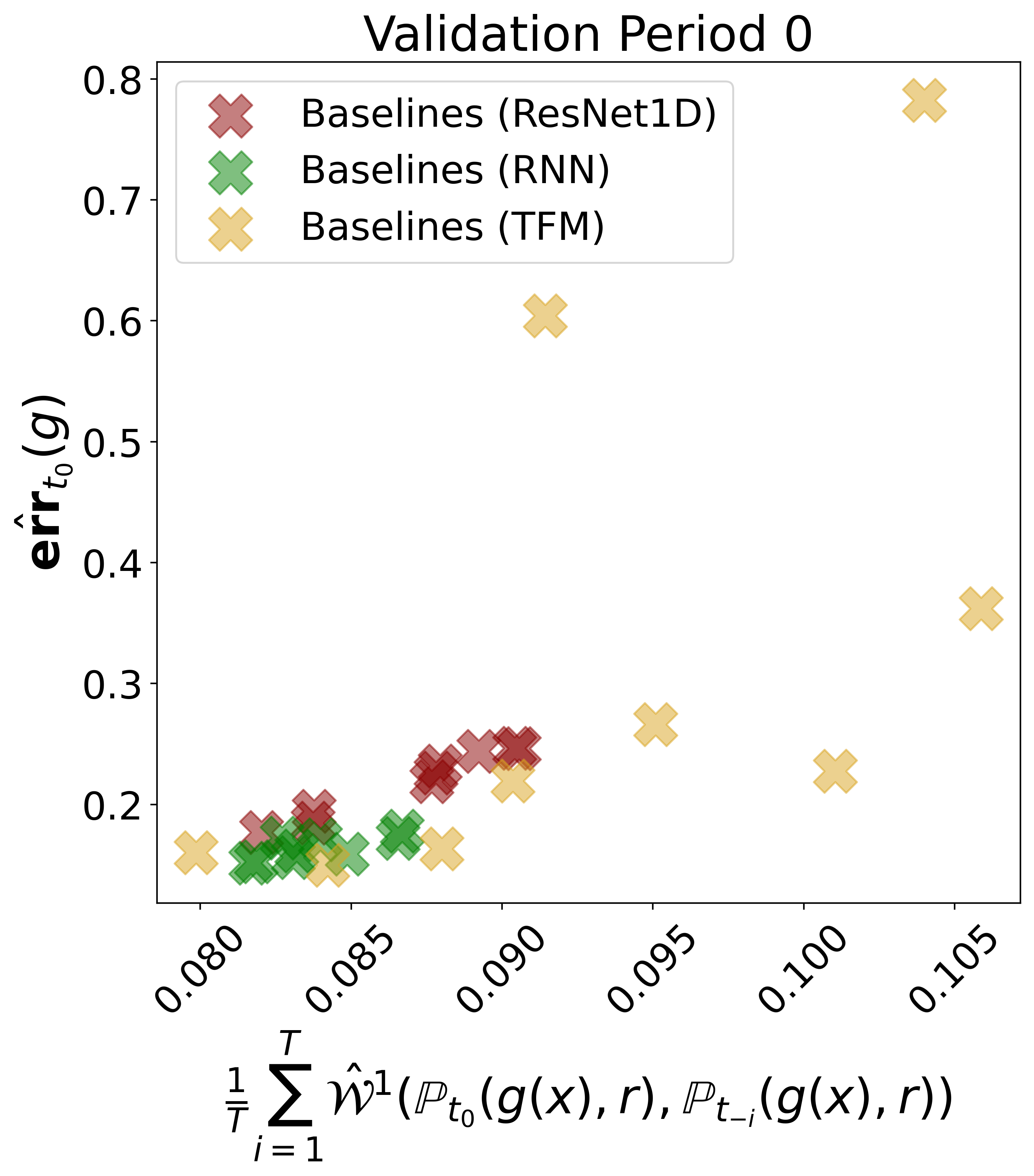}
        \includegraphics[width=0.32\linewidth]{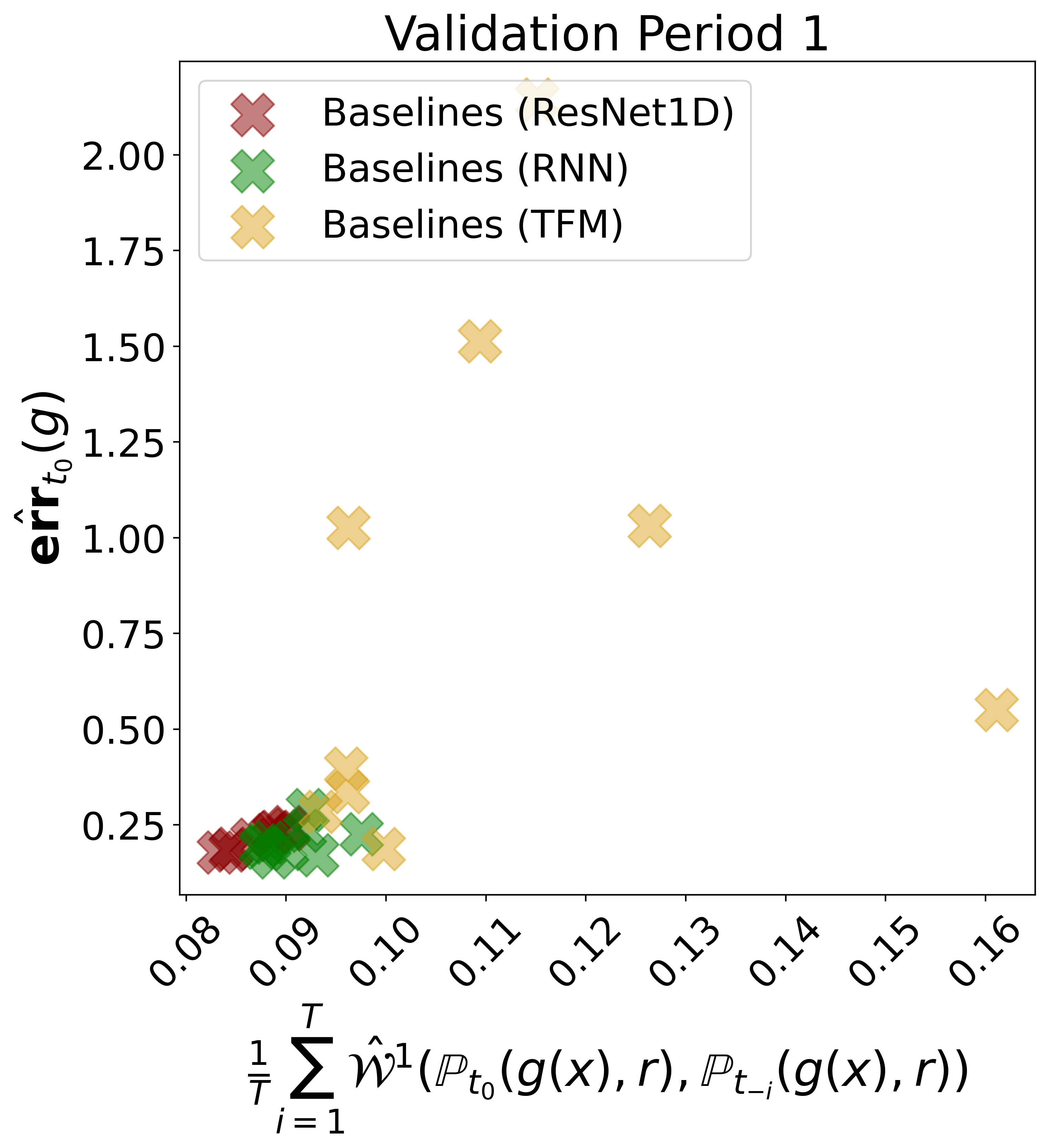}
        \includegraphics[width=0.32\linewidth]{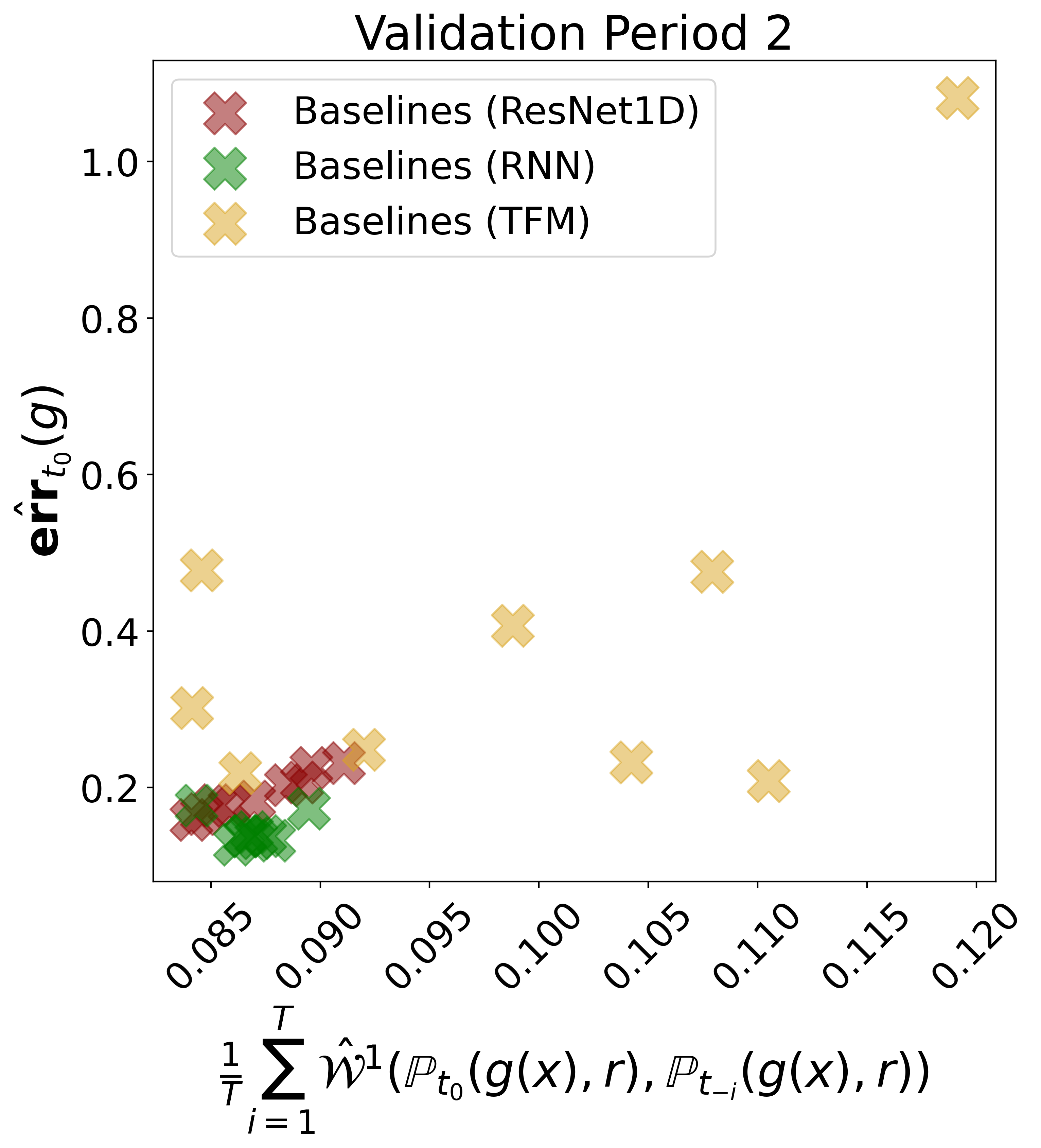}
        \label{fig:Wass_2b}
    }

    \caption{(a) The clusters of causal discovery algorithms over three validation periods. Supports induced by causal discovery algorithms are plotted together. The cases resulting from ResNet1D with $\tilde{K}=1$ are marked as squares. (b) The clusters of baseline algorithms over three validation periods. Supports induced by baseline algorithms are plotted together.}
    \label{fig:Wass_2}
\end{figure}

Notably, without causal discovery, TFM can result in significantly larger values of the Wasserstein term as well as OOS errors than the other architectures (Figure \ref{fig:Wass_2b}). This is likely due to stringent sample-size requirements or the inductive bias specifically associated with Transformer-like architectures, such as rank collapse \cite{dong2021attention}.\footnote{Rank collapse is a phenomenon where the learned representations of attention modules become overly concentrated in certain dimensions; This can exacerbates the issues of small datasets by effectively reducing the number of independent features the model learns, making it more prone to overfitting to noise.} In Figure \ref{fig:Wass_2a}, we zoom in on the causal discovery clusters. It is demonstrated that causal discovery can rescue the Transformer-like architecture, making the TFM supports comparable with others and even yielding lower OOS errors. This observation aligns with the largest discrepancy between TFM's baseline performances and their causal discovery counterparts as indicated in Table \ref{tab:performance_metrics}. Additionally, the support of ResNet1D clusters is wider in validation periods 0 and 2 (Figure \ref{fig:Wass_2a}), which can be resolved by enlarging $\tilde{K}$.

\begin{figure}[!t]
    \centering
    \subfloat[]{
        \includegraphics[width=0.32\linewidth]{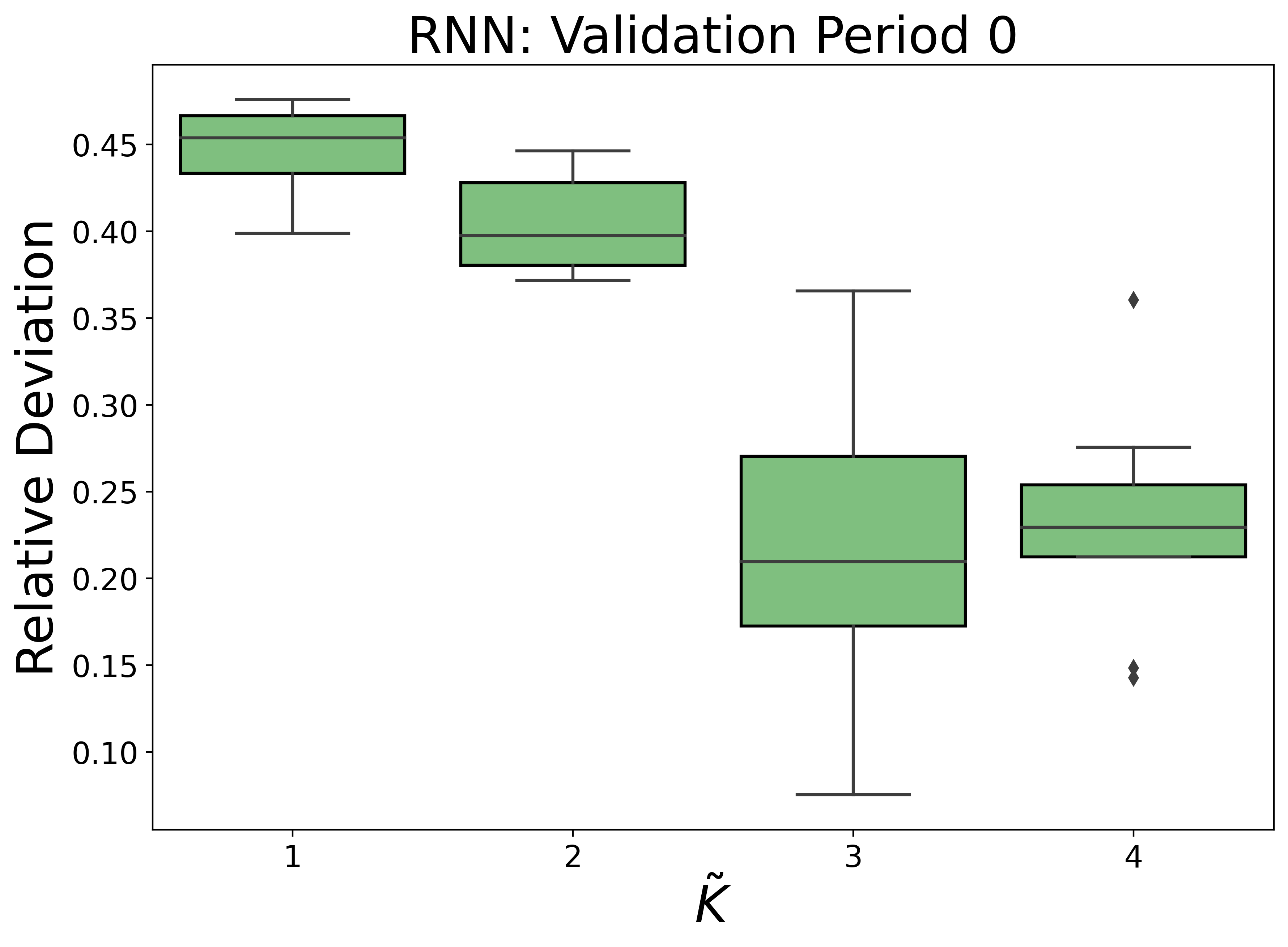}
    }
    \subfloat[]{
        \includegraphics[width=0.32\linewidth]{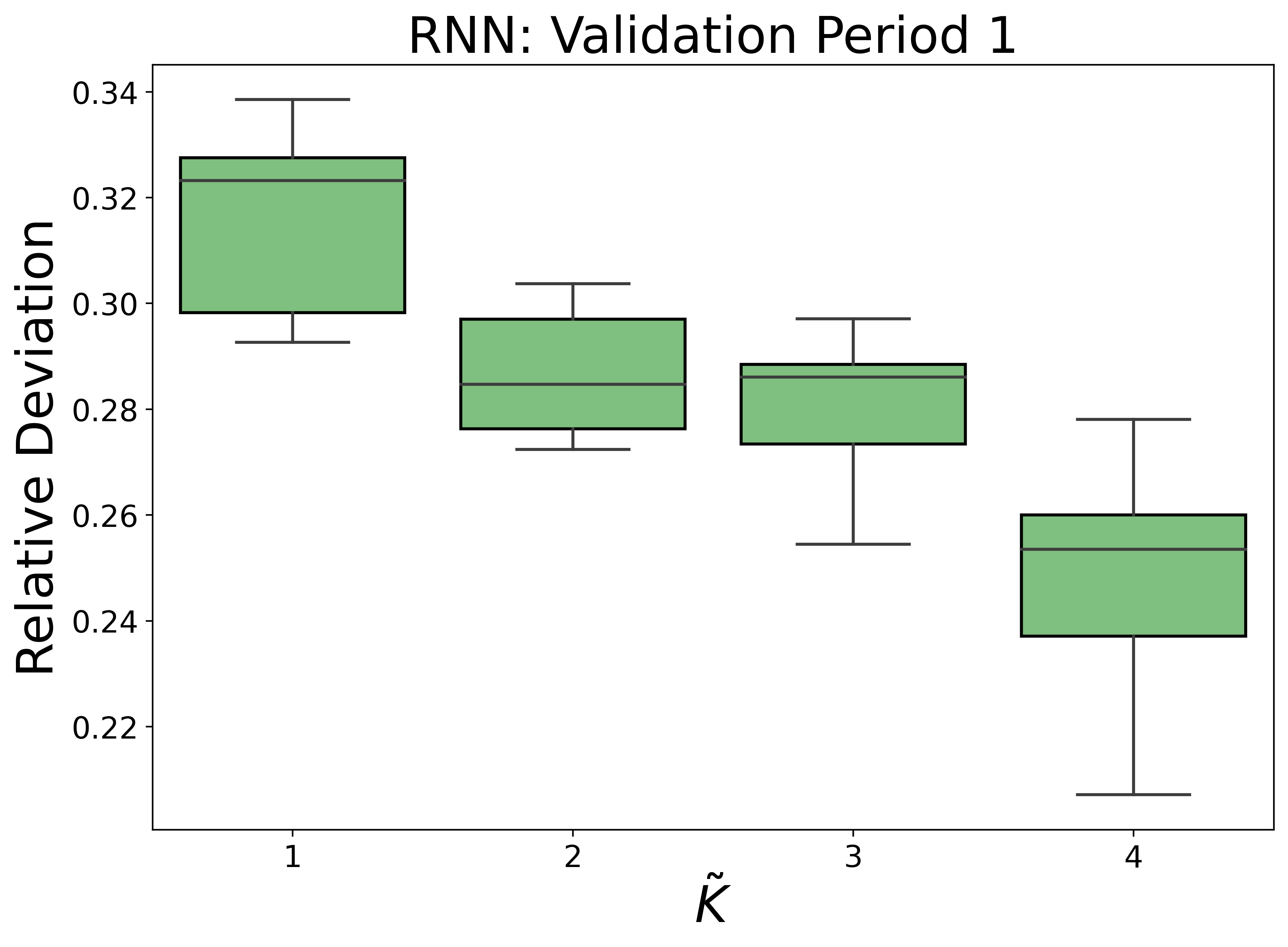}
    }
    \subfloat[]{
        \includegraphics[width=0.32\linewidth]{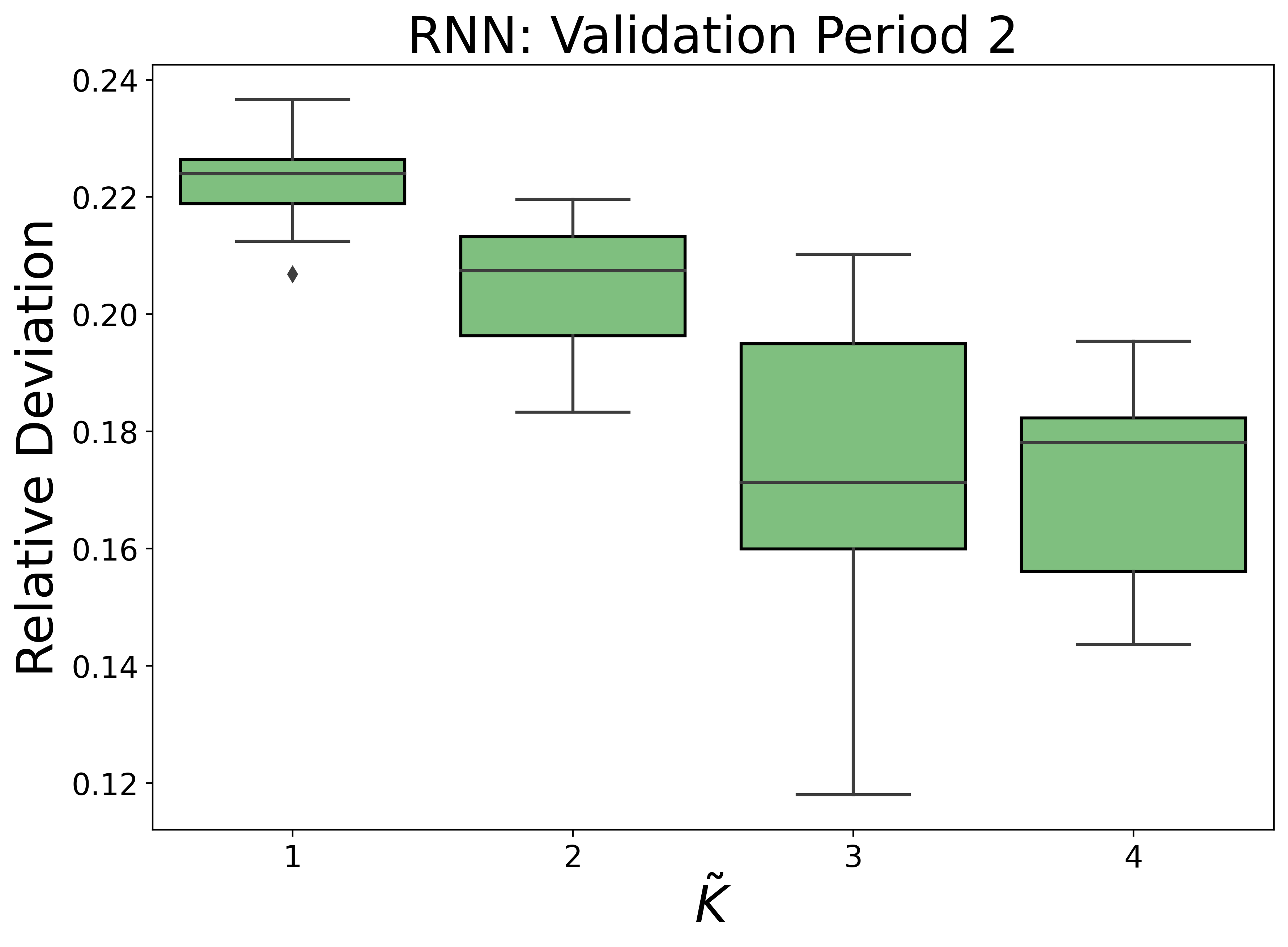}
    }\\
    \subfloat[]{
        \includegraphics[width=0.32\linewidth]{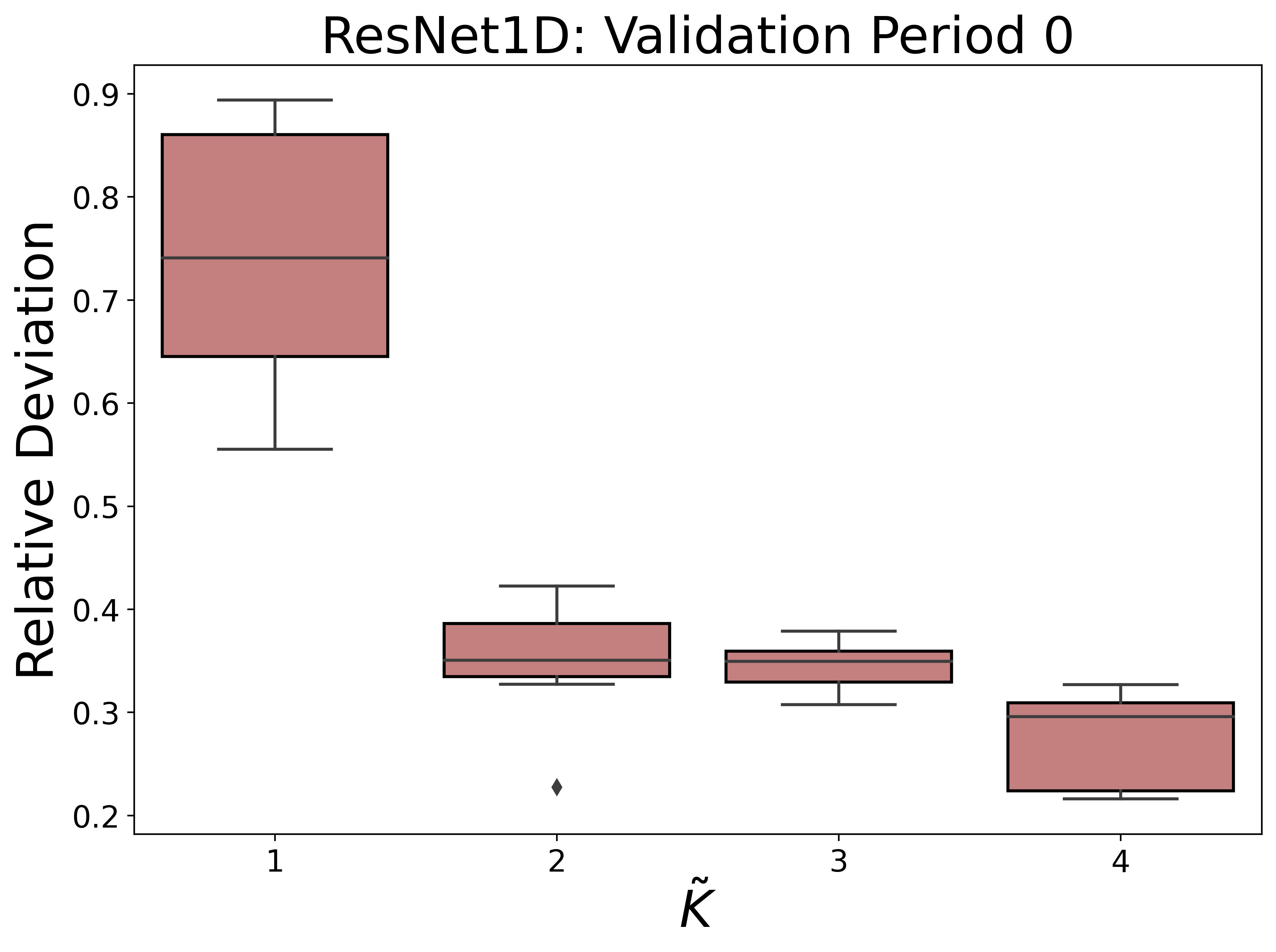}
    }
    \subfloat[]{
        \includegraphics[width=0.32\linewidth]{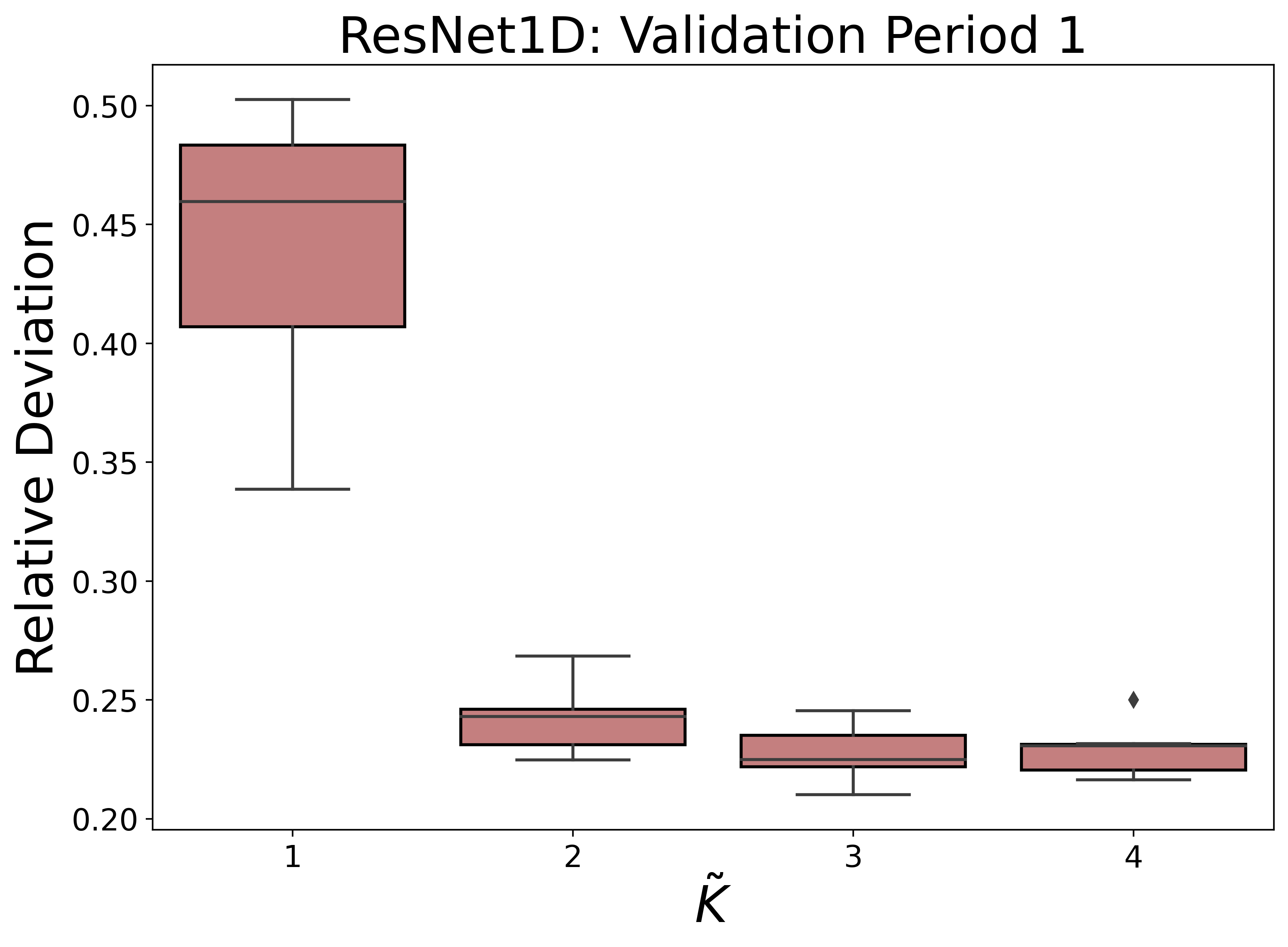}
    }
    \subfloat[]{
        \includegraphics[width=0.32\linewidth]{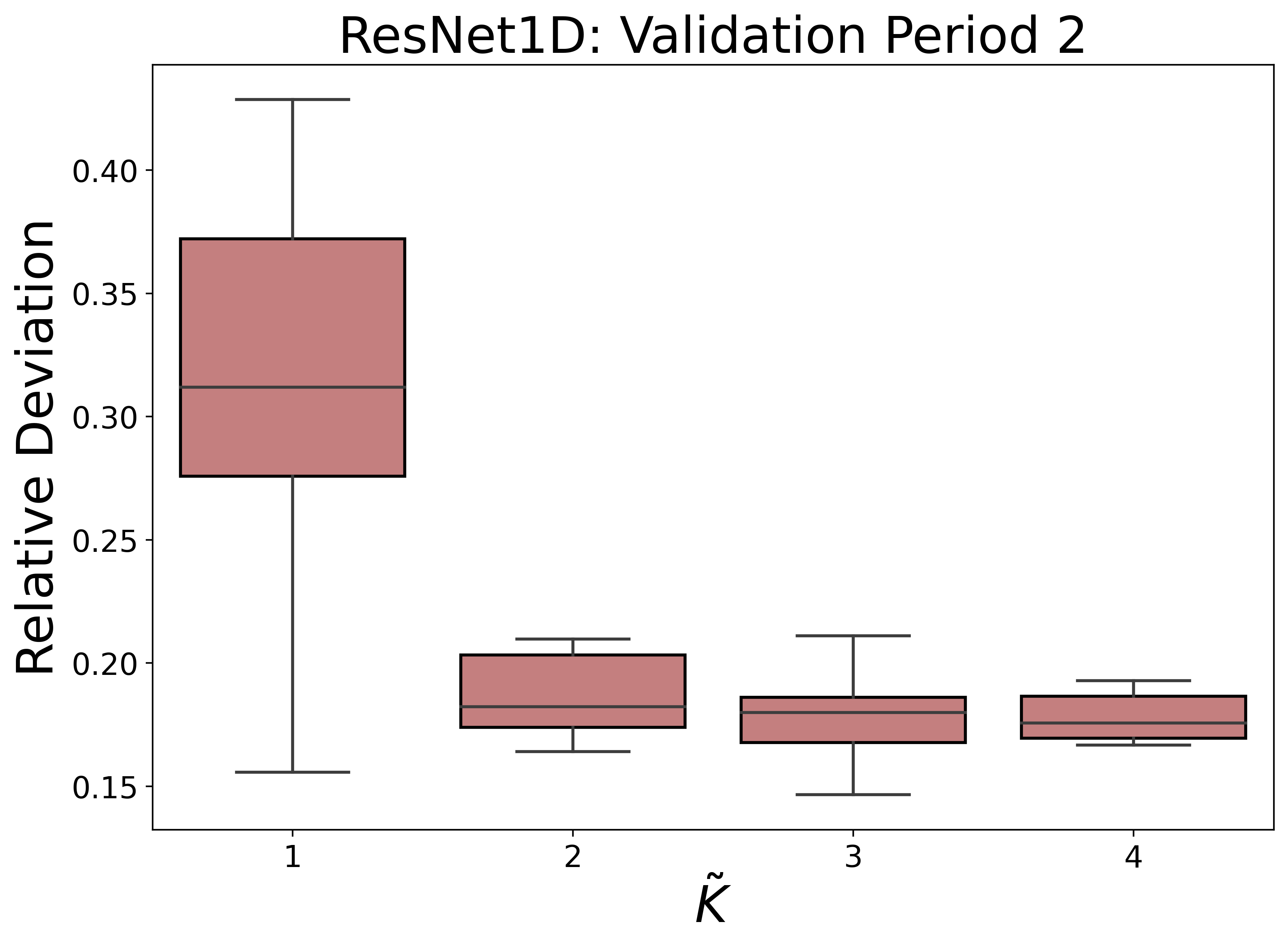}
    }\\

    \subfloat[]{
        \includegraphics[width=0.32\linewidth]{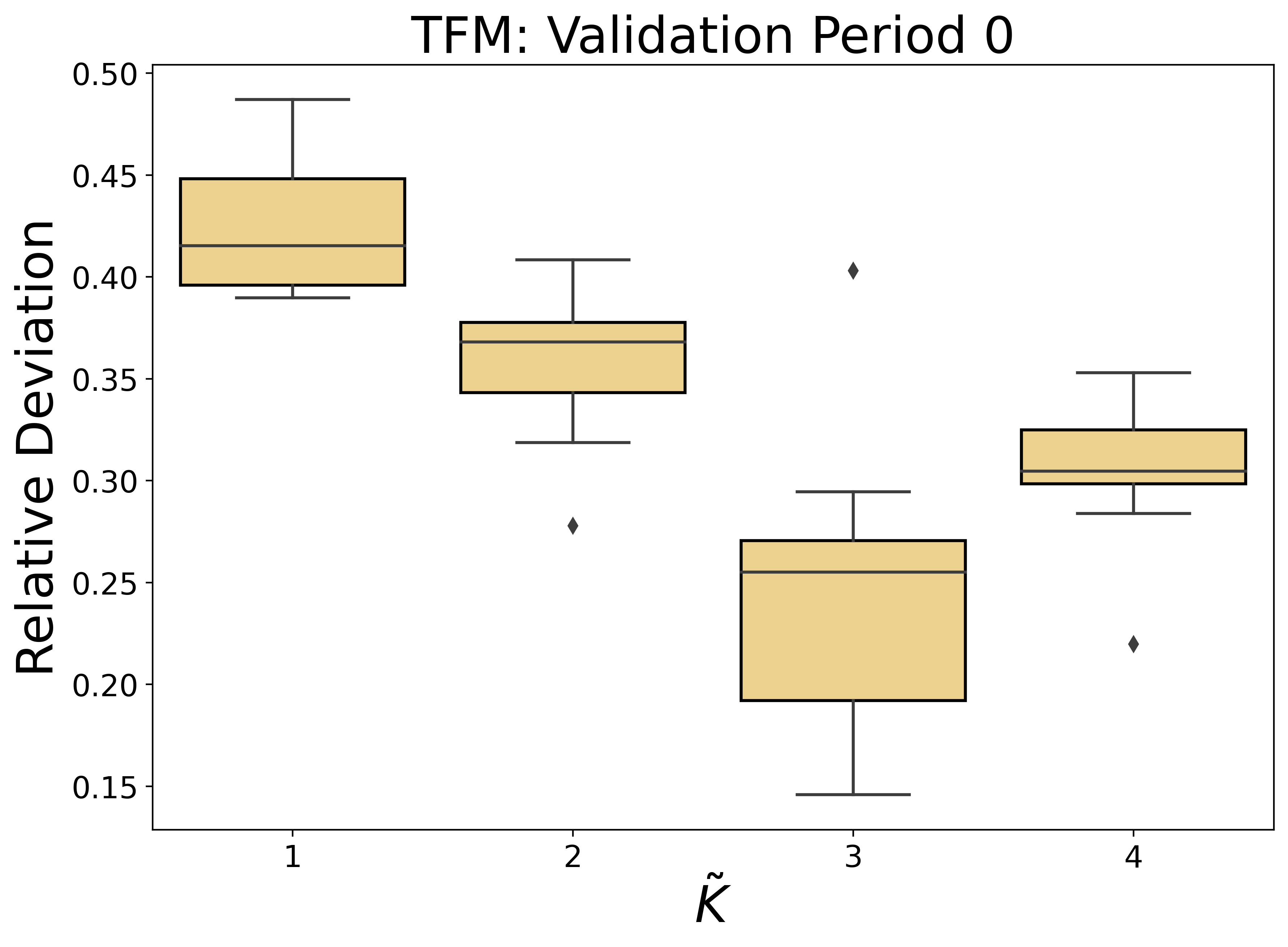}
    }
    \subfloat[]{
        \includegraphics[width=0.32\linewidth]{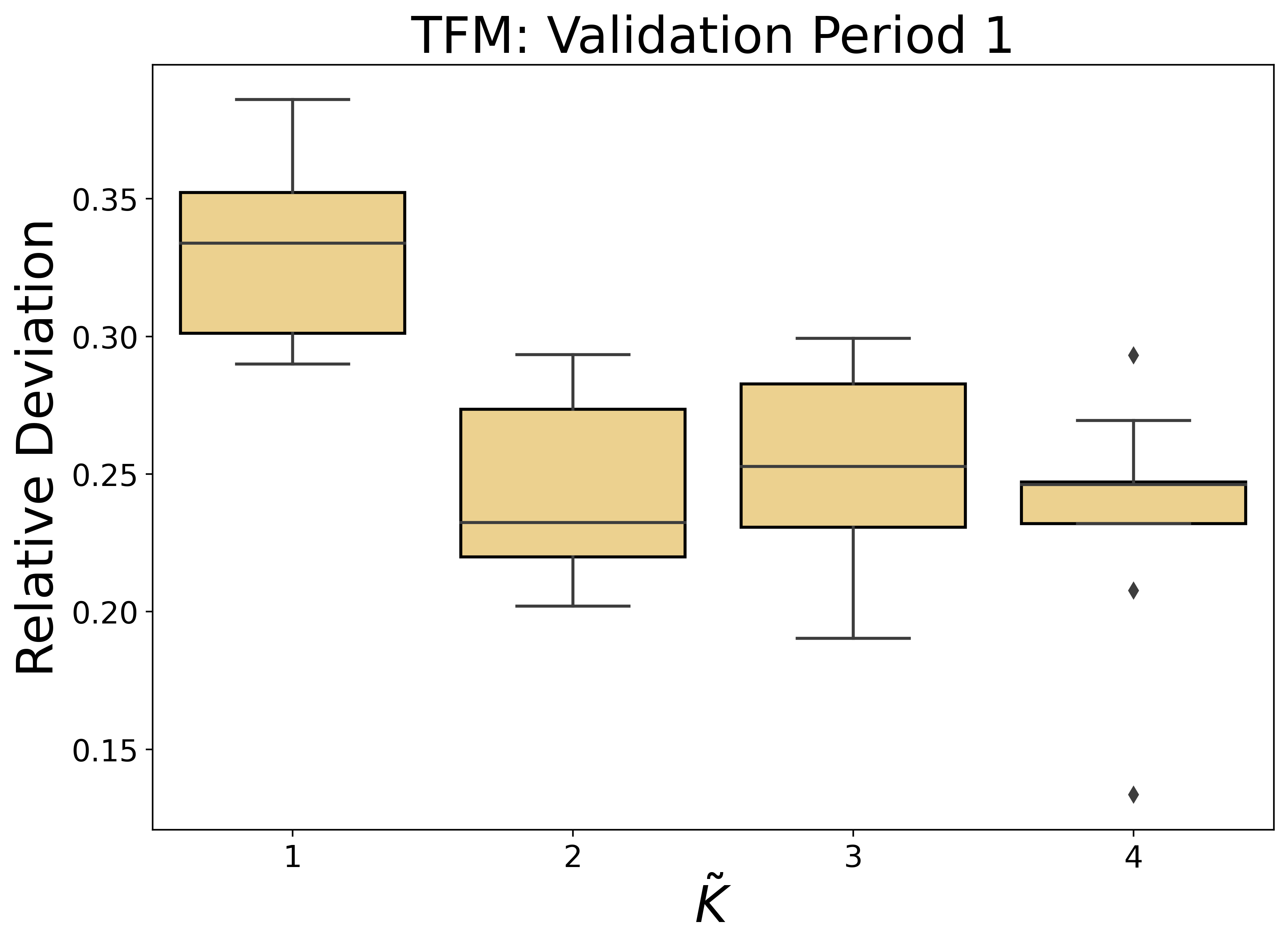}
    }
    \subfloat[]{
        \includegraphics[width=0.32\linewidth]{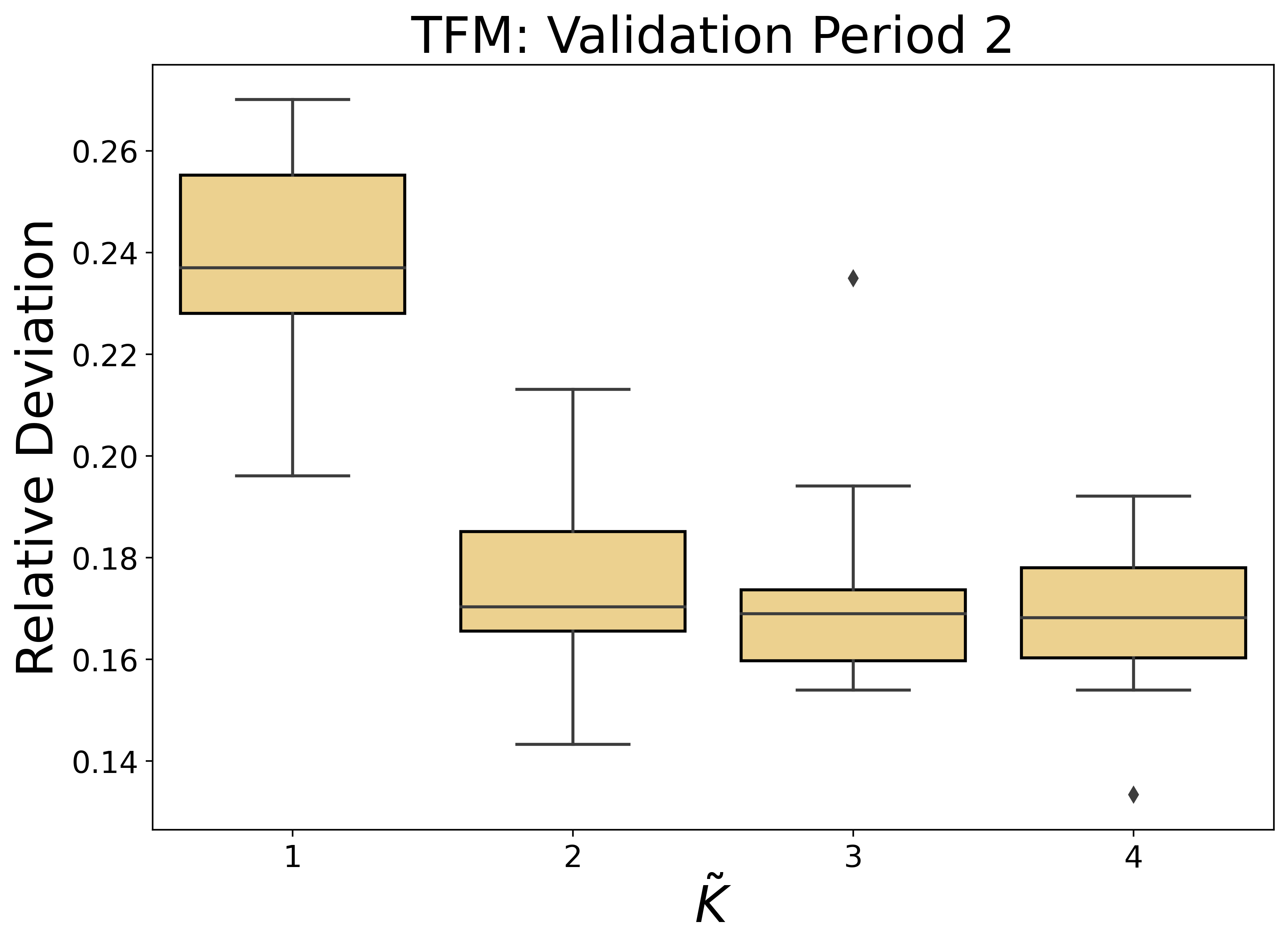}
    }

    \caption{Relative Deviation versus dimensionality $\tilde{K}$. Box plots are shown for 3 architectures (horizontal) $\times$ 3 validation periods (vertical). Each box plot depicts the distributions of Relative Deviation over 9 trained models for the cases $\tilde{K}=1$, 2, 3, and 4.}
    \label{fig:dims}
\end{figure}

\textbf{$\tilde{K}$ Comparisons.} We verify the role of dimensionality $\tilde{K}$ as indicated by Proposition \ref{prop:dim}, which argues that increasing $\tilde{K}$ allows the discovered causal features to attain lower residual levels without sacrificing invariance for the source domains. However, with a limited number of samples and overparameterized $\phi$, the estimations on the training set can be greatly affected by noise. A more plausible approach is to analyze the validation set, where the trend is less likely to be overwhelmed by noise. We freeze the learned $\phi$ from the training set and re-optimize $\bm{\nu} \in \mathbb{R}^{\tilde{K}}$ on the validation set. Then, we obtain the estimations of $\min_{\bm{\nu} \in \mathbb{R}^{\tilde{K}}} \mathcal{L}_{\text{pred}}(\bm{\nu}, \phi)$ and $\mathcal{L}_{\text{res}}(\phi)$ on the validation set. We consider the \textit{Relative Deviation} as the evaluation metric:
\begin{equation}
    \textit{Relative Deviation} := \frac{\hat{\mathcal{L}}_{\text{pred}}(\bm{\nu}, \phi) - \hat{\mathcal{L}}_{\text{res}}(\phi)}{\hat{\mathcal{L}}_{\text{res}}(\phi)}.
\end{equation}
As suggested by the proof of Proposition \ref{prop:Lpred} (see \textit{Supplementary Materials}), $\mathcal{L}_{\text{pred}}(\bm{\nu}, \phi) - \mathcal{L}_{\text{res}}(\phi)$ measures the deviation of causal coefficients. This deviation is normalized by dividing by the residual level since we are not interested in invariance with a high residual level (e.g., zero domain causal coefficients). Figure \ref{fig:dims} depicts the distributions of Relative Deviation over 9 trained models for the cases $\tilde{K}= 1$, 2, 3, and 4 across all architectures and validation periods. We observe a general pattern of decreasing Relative Deviation with increasing $\tilde{K}$. Notably, the reduction in Relative Deviation is significant from $\tilde{K}=1$ to $\tilde{K}=3$ in most cases. However, starting from $\tilde{K}=4$, the patterns are divergent. While some cases continue the downward trend, others show marginal improvements or even an increase in Relative Deviation. This may be due to the violated necessary condition as discussed in Proposition \ref{prop:dim} occurring around $\tilde{K}=4$. It may also result from the increased complexity of gradient computations with higher $\tilde{K}$, especially for the gradients with respect to $\mathcal{L}_{alig}(\phi)$. Note that Figure \ref{fig:ECs_dims} suggests $\tilde{K}=4$ is the optimal choice for the OOS domain, while Figure \ref{fig:dims} suggests $\tilde{K}=3$. This discrepancy may be explained by the difference between the validation period and the OOS period, where the 120-day model update frequency may be too large and could be causing such a discrepancy.

\textbf{Nonstationarity.} To investigate market nonstationarity, we simulate the samples of ideal joint errors corresponding to the random variable $J_T$. We randomly sample the OOS domain $t_0$ and then estimate $J_{T}(t_0) = \min_{g \in \Omega^{g}} \mathbf{err}_{t_0}(g) + \frac{1}{T}\sum_{i=1}^T \textbf{err}_{t_{-i}}(g)$ by neural network optimization to obtain one sample $\sim J_T$. We parameterize $\Omega^g$ by summing the outputs of the TFM, ResNet1D, and RNN architectures, and the training follows the baseline algorithms. To enhance readability, we make each source interval $(t_{-i-1}, t_{-i}]$ for $i=1, \cdots, T$ contain only one trading day so that $T$ domains correspond to $T$ trading days.\footnote{Note that the baseline algorithms are domain-irrelevant.} To ensure a sufficient number of samples, we let the OOS interval $(t_{-1}, t_0]$ contain 200 trading days.

Additionally, the training error may not be a suitable surrogate for the ideal joint error since overfitting the training noise can greatly underestimate the minima. Therefore, we divide the available samples into training and validation sets, and estimate the ideal joint error on the validation set using the prediction models trained on the training set. The validation set is formed by randomly picking 50 trading days from the OOS interval as well as $\frac{T}{4}$ trading days from the source domains, and the training set is constituted by the remaining trading days.

In our simulation, 32 OOS domains $t_0$ are randomly picked from the period between January 4, 2016, and March 4, 2024. For each $t_0$, $T$ varies in granularity of 100 from 200 to 1300. To reduce training randomness, for each $t_0$, we pretrain the model on the OOS domain and cease training if the validation MSE on the OOS domain does not decrease for five consecutive epochs. The pretrained models are then tuned for each $T$ with the sample importance of the OOS domain being upweighted by $\frac{T}{200}$, and again early-stopped according to joint MSE. Finally, we obtain the estimators by evaluating the tuned models on the validation sets. 

\begin{figure*}[!t]
    \centering
    \includegraphics[width=0.7\linewidth]{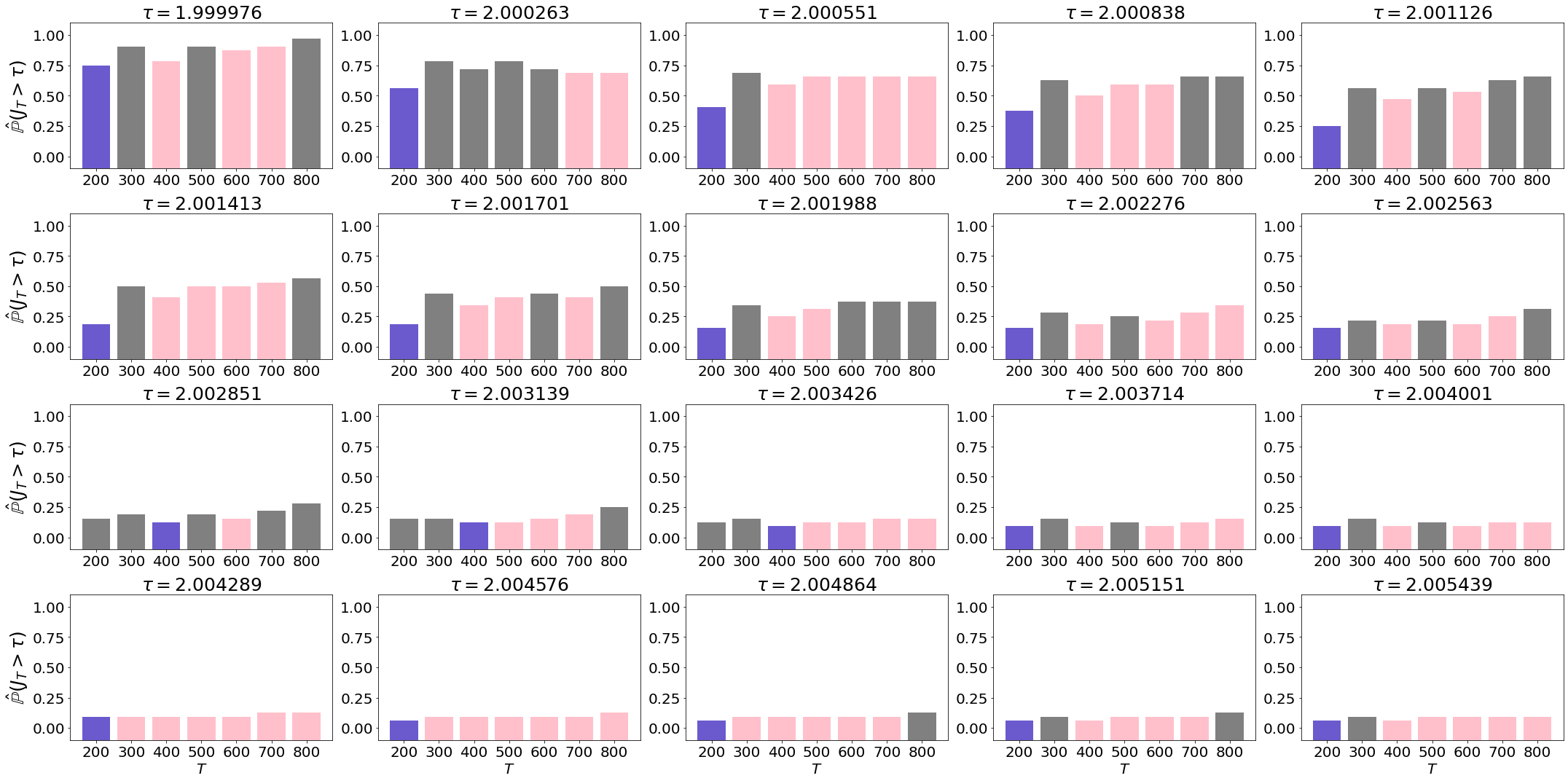}
    \caption{The estimated probabilities of joint errors surpassing 20 different $\tau$ values for $T \in \{200, 300, 400, 500, 600, 700, 800\}$. The global minima are colored in purple, and the $T$ values where the subsequent local minima are located are colored in pink (i.e., $T$ values that satisfy $\hat{\mathbb{P}}(J_T>\tau) \in \min\{\hat{\mathbb{P}}(J_{T'}>\tau) \mid T \leq T'\}$). Note that we exclude $T>800$ cases since they serve as buffers for the minimal estimation.}
    \label{fig:NAa}
\end{figure*}

Once we obtain the collection of estimations of ideal joint errors, we can estimate the $\mathbb{P}(J_T>\tau)$ for different choices of $\tau>0$. The $\tau$ values are obtained by 20 equally spaced numbers over the interval whose left and right ends are the 10\% and 90\% quantiles of the collection of estimations, respectively. Then, $\mathbb{P}(J_T>\tau)$ can be estimated by:
\begin{equation}
    \hat{\mathbb{P}}(J_T>\tau):= \frac{\text{\# of estimators $>\tau$}}{\text{Total \# of estimators}}.
\end{equation}
Figure \ref{fig:NAa} displays the estimated probabilities of joint errors exceeding 20 different thresholds ($\tau$) across a range of in-sample lengths $T \in \{200, 300, 400, 500, 600, 700, 800\}$. Each bar represents the estimated probability $\hat{\mathbb{P}}(J_T>\tau)$, with global minima highlighted in purple. These minima indicate the values of $T$ for which the joint error probability is lowest. Additionally, the values of $T$ where the subsequent local minima occur are marked in pink. These $T$'s are characterized by satisfying the condition $\hat{\mathbb{P}}(J_T>\tau) \in \min\{\hat{\mathbb{P}}(J_{T'}>\tau) \mid T \leq T'\}$. Cases where $T>800$ are excluded from this plot, as they are primarily used as buffers for estimating the minimal values. Notably, 18 out of 20 global minima occur at $T=200$, while two occur at $T=400$. Despite the exceptions, they all occur at the top-3 minimal choices of $T$, serving as strong evidence that picking a short-sized window preceding $t_0$ results in low $\lambda^*$ from the error bound (\ref{errorbound}). Moreover, the subsequent local minima are not far from $T$ where the global minima occur, and their locations show certain continuity. This supports the trade-off between the sample size and market nonstationarity.
\begin{figure}[t]
    \centering
    \includegraphics[width=0.7\linewidth]{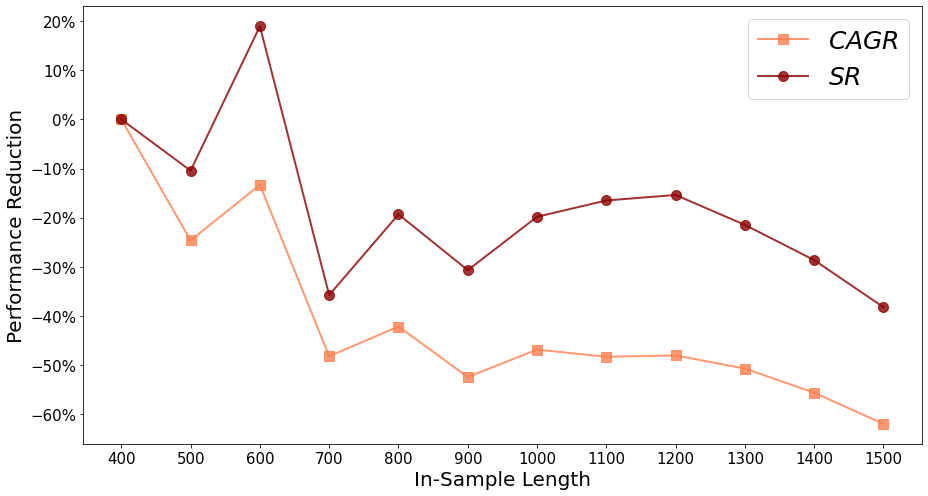}
    \caption{The performance reduction versus the number of trading days within the in-sample domain. The medians of CAGR and SR over 32 simulations are considered. The y-axis denotes the percentage of median reduction compared with the cases where the in-sample domain contains 400 trading days.}
    \label{fig:NAb}
\end{figure}

Next, we construct portfolios according to the signal outputs of the trained models used in estimating ideal joint errors. Specifically, for each sampled $t_0$, we construct the portfolios over the subsequent 120 trading days and obtain the CAGR and SR values. Figure \ref{fig:NAb} compares the medians of the CAGR and SR over 32 simulations with the number of trading days included in the training and validation sets. The y-axis represents the percentage reduction in median performance relative to the case where the in-sample domain includes 400 trading days. This visualization helps in understanding how the performance metrics change with varying in-sample lengths. To summarize, there is a downtrend for both SR and CAGR curves by increasing the number of samples, which is expected according to the observation that a larger $T$ causes higher risks of larger $\lambda^*$. Although larger sample numbers are desired, it seems large values of $\lambda^*$ can overwhelm the benefits of larger datasets. On the other hand, there is an exception at 600 in-sample length, where the median SR shows a 20\% improvement. Notably, 600 in-sample length corresponds to $T=400$, and this is where most second minima occur in Figure \ref{fig:NAa}, providing insights into the trade-offs between sample size and market nonstationarity.

\section{Conclusion}
In this study, we have established a novel linkage between factor models and Domain Generalization, providing extensive theoretical analysis on the generalization of prediction models as well as the impact of market nonstationarity. Although this represents a preliminary exploration in this direction, it lays the groundwork for more complex investigations. Future research could delve into Structural Causal Models (SCMs) with more intricate structures. For instance, one can incorporate an additional random variable to represent the market environment, potentially characterized by macroeconomic features. This may enhance the degree of common invariance between in-sample and out-of-sample (OOS) domains. Additionally, the source domains could be constituted beyond successive time steps. It is worth exploring Domain Discovery techniques that serve to enlarge $\Delta_{nc}-\Delta_c$ and therefore make causal features properly discovered. Lastly, our analysis is based on population objectives, whereas there is literature arguing that causal discovery techniques can perform differently on empirical objectives \cite{kamath2021does, rosenfeld2020risks}. More analysis is required to fill such a gap.

\bibliographystyle{IEEEtran}
\bibliography{main}

\end{document}